\documentclass{LMCS}

\def\dOi{9(3:19)2013}
\lmcsheading%
{\dOi}
{1--26}
{}
{}
{Oct.~12, 2012}
{Sep.~17, 2013}
{}

\usepackage{latexsym}

\usepackage{amssymb}

\usepackage{tikz}
\usepackage{hyperref,enumerate}
\usetikzlibrary{shapes.arrows,chains}

\newcommand{\CalL}{{\mathcal L}}
\newcommand{\range}{{\rm range}}
\newcommand{\bin}{{\rm bin}}

\theoremstyle{plain}
\theoremstyle{plain}
\theoremstyle{plain}
\theoremstyle{plain}
\theoremstyle{plain}
\theoremstyle{definition}\newtheorem{adefn}[thm]{Definition}
\theoremstyle{definition}
\theoremstyle{plain}
\theoremstyle{plain}

\def\niceqed{~~$\Box$}
\def\conv{{\rm{conv}}}
\def\sp{\\*\indent}

\begin{document}

\title[Automatic functions, linear time and learning]{Automatic
Functions, Linear Time and Learning\rsuper*}
\author[J.~Case]{John Case\rsuper a}%
\address{{\lsuper a}Department of Computer and Information Sciences,
University of Dela\-ware, Newark, DE 19716-2586, USA}
\email{case@cis.udel.edu}

\author[S.~Jain]{Sanjay Jain\rsuper b}%
\address{{\lsuper b}Department of Computer Science, National University
of Singapore, Singapore 117417, Republic of Singapore}
\email{sanjay@comp.nus.edu.sg}
\thanks{{\lsuper b}S.~Jain is supported in part by NUS grants C252-000-087-001 and
R252-000-420-112}

\author[S.~Seah]{Samuel Seah\rsuper c}%
\address{{\lsuper{c,d}}Department of Mathematics, National
University of Singapore, 10 Lower Kent Ridge Road,
Singapore 119076, Republic of Singapore}
\email{samseah85@yahoo.com and fstephan@comp.nus.edu.sg}

\author[F.~Stephan]{Frank Stephan\rsuper d}
\thanks{{\lsuper d}F.~Stephan is supported in part by NUS grant
R252-000-420-112.}

\begin{abstract}
\noindent
The present work determines the exact nature of {\em linear time computable\/}
notions which characterise automatic functions (those whose graphs are 
recognised by a finite automaton).
The paper also determines which type of
linear time notions permit full learnability for learning in the limit
of automatic classes (families of languages which are
uniformly recognised by a finite automaton). In particular it is shown
that a function is automatic
iff there is a one-tape Turing machine with a left end which computes the
function in linear time where the input before the computation and the
output after the computation both start at the left end. It is known
that learners realised as automatic update functions are restrictive for
learning. In the present work it is shown that one can overcome the problem
by providing work tapes additional to a resource-bounded base tape while
keeping the update-time to be linear in the length of the largest datum
seen so far.
In this model, one additional such work tape provides additional learning
power over the automatic learner model and two additional work tapes
give full learning power. Furthermore, one can also consider additional
queues or additional stacks in place of additional work tapes and for
these devices, one queue or two stacks are sufficient for full learning
power while one stack is insufficient.
\end{abstract}
\titlecomment{{\lsuper*}A preliminary version of this paper was presented at the conference
``Computability in Europe'' \cite{CJS12}.}



\ACMCCS{[{\bf Theory of computation}]: Models of
  computation---Abstract machines; Logic---Verification by model
  checking; Formal languages and automata theory; Theory and
  algorithms for application domains---Machine learning theory}

\maketitle
\vfill
\section{Introduction} \label{sec:introduction}

\noindent
In inductive inference, automatic learners and linear time learners have
played an important role, as both are considered as valid notions to model
severely resource-bounded learners.
On one hand, Pitt \cite{Pi89} observed that recursive learners can be made
to be linear time learners by delaying; on the other hand, when learners
are formalised by using automata updating a memory in each cycle with
an automatic function, the corresponding learners are not as powerful
as non-automatic learners \cite{JLS09} and cannot overcome their weakness
by delaying. The relation between these two models is that automatic learners
are indeed linear time learners \cite{CJLOSS11} but not vice versa.
This motivates to study the connection between linear time and automaticity
on a deeper level.
\sp
It is well known that a finite automaton recognises a regular language
in linear time. One can generalise the notion of automaticity from
sets to relations and functions \cite{Bl99,BG00,Ho76,Ho83,KN95,Ru08} 
and say that
a relation or a function is automatic iff an automaton recognises its graph,
that is, if it reads all inputs and outputs at the same speed and accepts
iff the inputs and outputs are related with each other, see
Section~\ref{se:autofuncchar} for a precise definition using the
notion of convolution. For automatic functions it is not directly
clear that they are in deterministic linear time, as recognising a graph
and {\em computing the output of a string from the input\/}
are two different tasks. Interestingly,
in Section~\ref{se:autofuncchar} below, it is shown that automatic functions
coincide with those computed by linear time one-tape Turing machines which
have the input and output both starting at the left end of the tape.
In other words, a function is automatic iff it is linear-time computable
with respect to the most restrictive variant of this notion;
increasing the number of tapes or not restricting the position of the
output on the tape results in a larger complexity class.
\sp
Section~\ref{se:lintimelearn} is dedicated to the question on how
powerful a linear time notion must be in order to capture full
learning power in inductive inference. For the reader's convenience,
here a short sketch of the underlying learning model is given:
Suppose $L \subseteq \Sigma^*$ is a language.
The learner gets as input a sequence $x_0, x_1, \ldots, $ of strings,
where each string in $L$ appears in the sequence and all the strings
in the sequence are from $L$ (such a sequence is called a text for $L$). 
As the learner is getting the input
strings, it conjectures a sequence of grammars $e_0,e_1,\ldots$
as its hypotheses about what the input language is.
These grammars correspond to some hypothesis space $\{H_e: e \in I\}$,
where $I$ is the set of possible indices and every possible learning
task equals to some $H_e$.
If this sequence of hypotheses converges to an index $e$ for the language $L$
(that is $H_e=L$),
then one can say that the learner has learnt the input language from
the given text. The learner learns a language $L$ if it learns it from all
texts for $L$. The learner learns a class $\CalL$ of languages if it learns
all languages from $\CalL$.
The above is essentially the model of learning in the limit proposed
by Gold \cite{Go67}. Equivalently, one can consider the learner as
operating in cycles, in $n$-th cycle it gets the datum $x_n$ and
conjectures the hypothesis $e_n$. In between the cycles, the learner
may remember its previous inputs/work via some memory. The
complexity of learners can be measured in terms of the complexity of
mapping the old memory and input datum to the new memory and hypotheses.
For automatic learners, one considers the above mapping to be given by
an automatic function.
\sp
In respect to the automatic learners
\cite{CJLOSS11,JLS09,JOPS10}, it has been the practice to study the
learnability of automatic classes (which are the set of all languages
in some automatic family) and furthermore only to permit hypothesis
spaces which are themselves automatic families containing the automatic
class to be learnt. It turned
out that certain automatic families which are learnable by a recursive
learner cannot be learnt by an automatic learner. The main weakness
of an automatic learner is that it fails to memorise all past data.
If one considers learning from fat text in which each datum occurs
infinitely often, then automatic learners have the same learning
power as recursive learners and their long-term memory can even
be restricted to the size of the longest datum seen so far \cite{JLS09},
the so called word size memory limitation.
\sp
Following the results of Section~\ref{se:autofuncchar},
one can simulate automatic learners by a learner using a one-tape Turing
machine which updates the content of the tape in linear time in
each round. In the present work this tape (called base tape)
is restricted in length by the length
of the longest datum seen so far --- as the corresponding word size
memory limitation of automatic learners studied in \cite{JLS09}. 
In each cycle, the
learner reads one datum about the set to be learnt and revises its
memory and conjecture. The question considered is how much extra power 
needs to be added to the learner for achieving full learnability; here
the extra power is formalised by permitting additional work tapes
which do not have length-restrictions; in each learning cycle the learner can,
however, only work on these tapes in time linear in the length of the
longest example seen so far. It can be shown using an archivation
technique, that two additional work tapes can store all the data observed in
a way that any learner can be simulated. When having only one
additional work tape, the current results are partial: using
a super-linear time-bound, one can simulate
any learner for a class consisting entirely of infinite languages;
furthermore, some classes not learnable by an automatic learner
can be learnt using one work tape. When considering additional
stacks in place of work tapes, two stacks are sufficient while one stack
gives some extra learning power beyond that of an automatic learner
but is insufficient to learn all in principle learnable classes.

\section{Automatic Functions and Linear Time} \label{se:autofuncchar}

\noindent
In the following, two concepts will be related to each other:
automatic functions and functions computed by position-faithful
one-tape Turing machines. In the following, a formal definition
of these two concepts is given. Automatic functions and structures
date back to the work of Hodgson \cite{Ho76,Ho83} and are based
on the concept of convolution. A convolution permits to write pairs
and tuples of strings by combining the symbols at the same position
to new symbols.

\begin{adefn}
\rm 
Let $\Sigma$ be a finite alphabet.
Let $\boxdot$ be a special symbol not in $\Sigma$.
The {\em convolution} $\conv(x,y)$ 
of two strings $x=x_1x_2\ldots x_m$ and $y=y_1y_2\ldots y_n$,
where $x_i,y_i \in \Sigma$,
is defined as follows. Let
$k=\max \{m,n\}$.
For $i \in \{1,2,\ldots,k\}$,
if $i \leq m$ then let $x'_i = x_i$ else let $x'_i = \boxdot$;
if $i \leq n$ then let $y'_i = y_i$ else let $y'_i = \boxdot$.
Now, the convolution is
$\conv(x,y)=(x'_1,y'_1) \linebreak[3] (x'_2,y'_2)
\linebreak[3] \ldots \linebreak[3] (x'_k,y'_k)$
where the symbols of this word are from the alphabet
$(\Sigma \cup \{\boxdot\}) \times (\Sigma \cup \{\boxdot\})$.
\end{adefn}

\noindent
Note that in the above definition, both $x$ and $y$ can be $\varepsilon$,
the empty string.
Similarly one can define the
convolution of a fixed number of strings. Now the convolution permits
to introduce automatic functions and relations.

\begin{adefn}[Hodgson \cite{Ho76,Ho83}]
\rm A function $f$, mapping strings to strings
(possibly over a different alphabet), is said to be 
{\em automatic} iff the set
$\{\conv(x,f(x)): x \in dom(f)\}$ is regular.

Similarly, an $n$-ary relation
$R \subseteq \{(x_1,x_2,\ldots,x_n): x_1,x_2,\ldots,x_n \in \Sigma^*\}$ is
automatic iff
$\{\conv(x_1,x_2,\ldots,x_n): (x_1,x_2,\ldots,x_n) \in R\}$ is
regular. An $n$-ary function $f$ is automatic iff
$\{\conv(x_1,x_2,\ldots,x_n,f(x_1,x_2,\ldots,x_n)): 
(x_1,x_2,\ldots,x_n) \in dom(f)\}$ is regular.
\end{adefn}

\noindent
Here a regular set \cite{HMU01} is a set which is recognised by a deterministic
finite automaton. This concept is equivalent to the one of sets
recognised by non-deterministic finite automata. Furthermore,
one can define regular sets inductively:
Every finite set of strings is regular. The concatenation of
two regular languages is regular, where $L \cdot H = \{xy: x \in L \wedge
y \in H\}$; similarly, the union, intersection and set difference of
two regular sets is regular. A further construct is the Kleene star,
$L^*$, of a regular language $L$ where $L^* = \{\varepsilon\} \cup L
\cup L \cdot L \cup L \cdot L \cdot L \cup \ldots = \{x_1 x_2 \ldots x_n:
x_1,x_2,\ldots,x_n \in L\}$. Note that
$L^*$ always contains the empty string $\varepsilon$. 
The above mentioned operations
are all that are needed, that is, a set is regular
iff it can be constructed from finite sets by using the above
mentioned operations in finitely many steps. 
The above inductive definition can be used
to denote regular sets by regular expressions which are written
representations of the above mentioned operations, for example,
$\Sigma^*$ represents the set of all strings over $\Sigma$ and
$\{00,01,10,11\}^* \cap (\{0,1\}^* \cdot \{0\} \cdot \{0,1\}^*)$
represents the set of all binary strings of even length which
contain at least one $0$.
\sp
The importance of the concept of automatic functions and automatic relations
is that every function or relation,
which is first-order definable from a finite number
of automatic functions and relations, is automatic again and the
corresponding automaton
can be computed effectively from the other automata. This gives the second
nice fact that structures consisting of automatic functions and relations
have a decidable first-order theory \cite{Ho83,KN95}.
\sp
A {\em position-faithful one-tape Turing machine} is a Turing
machine which uses a one-side infinite tape, with the left-end having
a special symbol $\boxplus$ which only occurs at this position and cannot
be modified. The input starts from the cell at the right 
of $\boxplus$ and is during the computation replaced by the output
which starts from the same cell. The end of input and output is the first
appearance of the symbol $\boxdot$ which is the default value of an empty
cell before it is touched by the head of the Turing machine or filled with
the input.

It is assumed that the Turing machine halts when it enters an accepting/final 
state (if ever).
A position-faithful one-tape Turing machine computes a function $f$, if
when started with tape content being $\boxplus$ $x$ $\boxdot^{\infty}$,
the head initially being at $\boxplus$, the Turing machine eventually
reaches an accepting state (and halts), 
with the tape content starting with $\boxplus f(x) \boxdot$.
Note that there is no restriction on the output beyond the
first appearance of $\boxdot$. Furthermore, a Turing machine can halt
without reaching an accepting state, in which case the computation is
not valid; this possibility is needed when a non-deterministic Turing
machine has to satisfy a time bound on the duration of the computation.
\sp
Though the requirement of ``position-faithfullness'' seems to be a bit
artificial, it turns out that it is a necessary requirement. This is
not surprising, as moving $i$ bits by $j$ cells requires, in the worst case,
proportional to $i \cdot j$ steps. So sacrificing the requirement
of position-faithfullness clearly increases the class. For example,
the function which outputs the binary symbols between the first and
second occurrence of a digit other than $0$ and $1$ of an input would
become linear time computable by an ordinary one-tape Turing machine
although this function is not linear time
computable by a position-faithful one-tape Turing machine. Such an
additional side-way to move information (which cannot
be done in linear time on one-tape machines) 
has therefore been excluded from the model.
The functions computed by position-faithful one-tape Turing
machines are in a certain sense a small natural class of linear time
computable functions.
\sp
Some examples of automatic functions are those which append to or delete in
a string some characters as long as the number of these characters is bounded
by a constant. For example a function deleting the first 
occurrence (if any)
of $0$ in a string would be automatic; however, a function deleting all
occurrences of $0$ is not automatic. Below is a more comprehensive example
of an automatic function.

\begin{exa} \label{ex:firstlastexchange}
Suppose $\Sigma = \{0,1,2\}$. Suppose $f$ is a mapping from $\Sigma^*$ to
$\Sigma^*$ such that $f(x)$ interchanges the first and last symbol in $x$;
$f(\varepsilon) = \varepsilon$.
Then $f$ is automatic and furthermore $f$ can also be computed by a
position-faithful one-tape Turing machine.
\sp
To see the first, note that the union of the set
$\{\varepsilon\} \cup \{(a,a): a \in \Sigma\}$ and
all sets of the form $\{(a,b)\} \cdot \{(0,0),(1,1),(2,2)\}^* \cdot
\{(b,a)\}$ with $a,b \in \Sigma$ is a regular set.
Thus $\{\conv(x,y): x \in \Sigma^* \wedge y = f(x)\}$
is a regular set and $f$ is automatic.
\sp
A position-faithful one-tape Turing machine would start on the
starting symbol $\boxplus$ and go one step right. In the case
that there is a $\boxdot$ in that cell,
the machine halts. Otherwise it memorises in
its state the symbol $a$ there. Then it goes right until it finds
$\boxdot$; it then goes one step left. 
The Turing machine then memorises the symbol $b$ at this position
and replaces it by $a$. It then goes left until it finds $\boxplus$,
goes one step right and writes $b$.
\end{exa}

\noindent
That $f$ in the preceding example can be computed in both ways
is not surprising, but indeed a consequence of the main result
of this section which states that the following three
models are equivalent:
\begin{iteMize}{$\bullet$}
\item automatic functions;
\item functions computed in deterministic linear time by a position-faithful
      one-tape Turing machine;
\item functions computed in non-deterministic linear time by a
      position-faithful one-tape Turing machine.
\end{iteMize}
This equivalence is shown in the following two results, where the first
one generalises prior work \cite[Remark 2]{CJLOSS11}.

\begin{thm} \label{th:islinear}
Let $f$ be an automatic function. Then there is a deterministic linear time
one-tape position-faithful Turing machine which computes $f$.
\end{thm}

\proof
The idea of the proof is to simulate the behaviour of a deterministic
finite automaton recognising the graph of $f$. The Turing Machine
goes from the left to the right over the input word and takes note of which
states of the automaton can be reached from the input with only one unique 
possible output.
Once the automaton reaches an accepting state in this simulation
(for input/output pairs), the simulating Turing machine turns back (that
is, it goes from right to left over the tape) 
converting the sequence of inputs and the stored information about 
states as above into that output which produces the unique accepting run 
on the input. Now the formal proof is given.
\sp
Suppose that a deterministic automaton with $c$ states (numbered $1$
to $c$, where $1$ is the starting state) accepts a word
of the form $\conv(x,y) \cdot (\boxdot,\boxdot)$ iff $x$
is in the domain of $f$ and $y = f(x)$; the automaton rejects
any other sequence. Note that this small modification of the
way the convolution is represented simplifies the proof.
As $f(x)$ depends uniquely on $x$, any string of the form
$\conv(x,y) \cdot (\boxdot,\boxdot)$ 
accepted by the automaton satisfies $|y| \leq |x|+c$. 
Let $\delta$ be the transition function for the automaton above
and $\hat{\delta}$ be the corresponding extended transition
function~\cite{HMU01}.
\sp
Suppose that the input is $x=x_1x_2\ldots x_r$.
Let the cell number $k$ be that cell which carries the input $x_k$ 
(with cell $0$ carrying $\boxplus$), that is the $k$-th cell to the
right of $\boxplus$; $\boxplus$ is in cell number $0$. 
Note that the Turing Machine described below
does not use the cell number in its computation; 
the numbering is used just for ease of notation.
The simulating Turing machine
uses a larger tape alphabet containing extra symbols from
$(\Sigma \cup \boxdot) \times \{+,-,*\}^c$,
that is, one considers the additional symbols consisting of tuples
of the form $(a,s_1,s_2,\ldots,s_c)$, where $a \in \Sigma \cup \{\boxdot\}$ 
and
$s_i \in \{-,+,*\}$. These symbols are written temporarily onto
the tape while processing the word from the left to the right and later
replaced when coming back from the right to the left.
\sp
Intuitively, during the computation while going from left to right, 
for cell number $k$,
one wishes to replace $x_k$ by the tuple $(x_k,s^k_1,s^k_2,\ldots,s^k_c)$ where,
for $d \in \{1,2,\ldots,c\}$: $s^k_d = -$ iff there is no word
of the form $y_1y_2\ldots y_{k-1}$ such that the automaton on input
$(x_1,y_1) \linebreak[3] (x_2,y_2) \linebreak[3] \ldots \linebreak[3]
(x_{k-1},y_{k-1})$ reaches the state $d$
(that is, for no $y_1y_2\ldots y_{k-1}$, 
$\hat{\delta}(1,(x_1,y_1) (x_2,y_2) \ldots
(x_{k-1},\linebreak[3] y_{k-1}))=d$); 
$s^k_d = +$ iff there is exactly one
such word; $s^k_d = *$ iff there are at least two such words. Here
the $x_i$ and $y_i$ can also be $\boxdot$ (when $i$ is larger than
the length of the relevant string, for example $x_i=\boxdot$ for $i>r$).
\sp
For doing the above, the Turing machine simulating the automaton replaces the 
cell to the right of $\boxplus$, that is,
the cell containing $x_1$, by $(x_1,+,-,\ldots,-)$.
Then, for the $k$-th cell, $k>1$, 
to the right of $\boxplus$,
with entry $x_k$ (from the input or
$\boxdot$ if $k>r$) the Turing machine replaces $xs_k$ by
$(x_k,s^k_1,s^k_2,\ldots,s^k_c)$ under the following conditions,
(where the entry in the cell to the left was $(x_{k-1},s^{k-1}_1,s^{k-1}_2,
\ldots,s^{k-1}_c)$ and where $d$ ranges over $\{1,2,\ldots,c\}$): 
\begin{iteMize}{$\bullet$}
\item $s^k_d$ is $+$ iff there is exactly one $(y_{k-1},d')
  \in (\Sigma \cup \{\boxdot\}) \times \{1,2,\ldots,c\}$ 
  such that $s^{k-1}_{d'}$ is $+$ and $\delta(d',(x_{k-1},y_{k-1}))=d$ 
  and there is no pair $(y_{k-1},d')
  \in (\Sigma \cup \{\boxdot\}) \times \{1,2,\ldots,c\}$ 
   such that $s^{k-1}_{d'}$ 
  is $*$ and $\delta(d',(x_{k-1},y_{k-1}))=d$;
\item $s^k_d$ is $*$ iff there are at least two pairs $(y_{k-1},d')
  \in (\Sigma \cup \{\boxdot\}) \times \{1,2,\ldots,c\}$ 
   such that
  $s^{k-1}_{d'}$ is $+$ and $\delta(d', (x_{k-1},y_{k-1}))=d$
  or there is at least one pair $(y_{k-1},d')
  \in (\Sigma \cup \{\boxdot\}) \times \{1,2,\ldots,c\}$ 
  such that $s^{k-1}_{d'}$
  is $*$ and $\delta(d',(x_{k-1},y_{k-1}))=d$;
\item $s^k_d$ is $-$ iff for all pairs $(y_{k-1},d')
  \in (\Sigma \cup \{\boxdot\}) \times \{1,2,\ldots,c\}$ 
  such that 
   $\delta(d',(x_{k-1},\linebreak[3] y_{k-1}))=d$, it holds that 
  $s^{k-1}_{d'}$ is $-$.
\end{iteMize}
Note that the third case applies iff the first two do not apply. The automaton
replaces each symbol in the input as above until it reaches the cell 
where the intended symbol $(x_k,s^k_1,s^k_2,\ldots,s^k_c)$ has
$s^k_d = +$ for some accepting state $d$.
(Note that the accepting states occur in the automaton only if both
the input and output are exhausted by the convention made above.)
If this happens, the Turing machine turns around, memorises
the state $d$, erases this cell (that is, writes $\boxdot$)
and goes left. 
\sp
When the Turing machine moves left from the cell number $k+1$ to the cell 
number $k$ (which contains the entry $(x_k,s^k_1,s^k_2,\ldots,s^k_c)$),
where the state memorised for the cell number $k+1$ is $d'$, 
then it determines the
unique $(d,y_k) \in \{1,2,\ldots,c\} \times
 (\Sigma \cup \{\boxdot\})$ such that $s^k_d = +$ and
$\delta(d,(x_k,y_k))=d'$;
then the Turing machine 
replaces the symbol on cell $k$ by $y_k$.
Then the automaton keeps the state $d$ in the memory and goes to
the left and repeats this
process until it reaches the cell which has the symbol
$\boxplus$ on it. Once the Turing machine reaches there, it
terminates.
\sp
For the verification, note that the output $y=y_1y_2\ldots$ 
(with $\boxdot$ appended) satisfies
that the automaton, after reading $(x_1,y_1) \linebreak[3] (x_2,y_2)
\linebreak[3] \ldots \linebreak[3] (x_k,y_k)$, is always in a state $d$
with $s^{k+1}_d = +$,
as the function value $y$ is unique in $x$;
thus, whenever the automaton ends up in an accepting state $d$ with
$s^{k+1}_d = +$ then the input-output-pair
$\conv(x,y) \cdot (\boxdot,\boxdot)$ has
been completely processed and $x \in dom(f) \wedge f(x)=y$ has been
verified. Therefore, the Turing machine can turn and follow the
unique path, marked by $+$ symbols, backwards in order to reconstruct
the output from the input and the markings. All superfluous symbols
and markings are removed from the tape in this process. 
As the automaton accepts $\conv(x,y) \cdot (\boxdot,\boxdot)$,
and $y$ depends uniquely on $x$, $|y| \leq |x|+c$. 
Hence the runtime
of the Turing machine is bounded by $2 \cdot (|x|+c+2)$, that is, the
runtime is linear.\niceqed

\begin{rem}
Note that the Turing machine in the above theorem 
makes two passes, one from the origin to
the end of the word (plus maybe constantly many more cells)
and one back. These two passes are needed
for a deterministic Turing machine: Recall
the function $f$ from Example~\ref{ex:firstlastexchange} with
$f(x_1 x_2 \ldots \linebreak[3] x_{k-1} x_k) = x_k \, x_2 \ldots x_{k-1} x_1$
for all non-empty words $x_1 x_2 \ldots x_{k-1} x_k$. When starting at the left
end, the machine has first to proceed to the right end to read the last
symbol before it can come back to the left end in order to write that
symbol into the new first position. Hence the runtime of the one-tape
deterministic Turing machine (for the simulation as in
Theorem~\ref{th:islinear})
cannot be below $2 \cdot |x|$ for an input $x$. Non-deterministic Turing
machines can, however, perform this task with one pass.
\end{rem}

\noindent
For the converse direction of the equivalence of the two models
of computation, that is, of automatic functions and position-faithful linear
time Turing machines, assume that a function is computed by a position-faithful
non-deterministic one-tape Turing machine in linear time.
For an input $x$, any two non-deterministic
accepting computations have to produce the same output $f(x)$. Furthermore,
the runtime of each computation has to follow the same linear bound
$c \cdot (|x|+1)$, independent of whether the computation ends up in an
accepting state or a rejecting state.

\begin{thm} \label{th:isautomatic}
Let $f$ be a function computed by a non-deterministic one-tape
position-faithful Turing machine in linear time.
Then $f$ is automatic.
\end{thm}

\proof
The proof is based on crossing-sequence methods, see \cite{He65,He65b}
and \cite[Section VIII.1]{Od99}. The idea is to show that $f$ is automatic
by providing a non-deterministic automaton
which recognises the graph $f$ of the function by going
symbol by symbol over the convolution of input and output and for each
symbol, the automaton guesses, for the Turing Machine on the corresponding
input, the crossing sequence on the right side
and verifies that this crossing-sequence is
compatible with the previously guessed crossing-sequence on the
left side of the symbol plus the local transformation of the respective
input symbol to the output symbol. This is now explained in more detail.
\sp
Without loss of generality one can assume that the position-faithful
Turing machine $M$ computing $f$ starts at $\boxplus$
and returns to that position at the end; a computation accepts only when the
full computation has been accomplished and the automaton has returned to
$\boxplus$. By a result of Hartmanis~\cite{Ha68} and
Trakhtenbrot~\cite{Tr64}, there is a constant $c'$ such that an accepting
computation visits each cell of the tape at most $c'$ times; otherwise
the function $f$ would not be linear time computable. This permits to
represent the computation locally by considering for each visit to a cell --- 
the direction from which the Turing machine $M$ entered the cell, in which 
state it was, what activity it did and in which direction it left the cell. 
Below, the $k$-th cell to the right of $\boxplus$ is referred to as cell
number $k$.
The local computation at the cell number $k$
can be considered as a tuple
$(x_k,is^1_k,os_k^1,d_k^1,z_k^1, is_k^2,os_k^2,d_k^2,z_k^2, \ldots, 
is_k^{r_k},os_k^{r_k},d_k^{r_k},z_k^{r_k})$, for some $r_k \leq c'$,
where $x_k$ is the initial symbol at the cell number $k$, and for each $j$,
$is_k^j$ denotes the state the Turing machine $M$ was in
when it visited the cell number $k$ for the $j$-th time,
$os_k^j$ is the state the Turing machine $M$ was in when it 
left the cell number $k$ after the $j$-th visit,
$d_k^j$ is the direction in which the 
Turing machine $M$ left after the $j$-th visit,
and $z_k^j$ is the symbol written in the cell number $k$ 
by the Turing machine $M$
during the $j$-th visit;
$r_k$ here denotes the total number of visits of the Turing machine $M$
to the $k$-th cell. Note that the number of possibilities for
the local computation as above is bounded by a constant.
\sp
As an intermediate step one shows that
a non-deterministic finite state automaton can recognise the set
\begin{quote}
  $A = \{\conv(x\cdot \boxdot^s,y\cdot \boxdot^{s+|x|-|y|}): x \in dom(f)
  \wedge y = f(x) \wedge s>0 \wedge s+|x| > |y| \wedge$
  the Turing machine $M$ on input $x$
  does not move beyond cell number $s-1+|x|\}$.
\end{quote}
This is done by initially guessing the local computation at $\boxplus$
(the $0$-th cell).
Then on each subsequent input $(x_k,y_k)$ (where $x_k$ or $y_k$
might be $\boxdot$), starting with $k=1$,
the automaton (i) guesses the local computation at the $k$-th cell, 
(ii) checks that this guess in (i) is consistent with the local
computation guessed
at cell $k-1$ (that is, each time the Turing machine $M$ moved
from cell $k-1$ to cell $k$ or cell $k$ to $k-1$, the corresponding guessed
leaving/entering states match), (iii) the computation within the cell
is consistent with the
Turing machine $M$'s state table (that is, either each of the entries
$is_k^j,os_k^j,d_k^j,z_k^j$ satisfies that Turing machine has transition
from state $is_k^j$ on reading input $z_k^{j-1}$ to
state $os_k^j$ writing $z_k^j$ in the cell and moving in direction
$d_k^j$, where $z_k^0=x_k$ and $z_k^{r_k}=y_k$ or the Turing machine
does not reach this cell and $y_k = x_k$), (iv) 
for the last input the automaton also checks that it is of the form
$(\boxdot,\boxdot)$ and that the Turing machine $M$ does not reach
this cell.
\sp
If at the end, all the computation and guesses are consistent then the
automaton accepts.
The automaton thus passes over the full word and
accepts $\conv(x\cdot \boxdot^s,y\cdot \boxdot^{s+|x|-|y|})$ iff the 
non-deterministic computation transforms 
$\boxplus x \boxdot^s$ into 
$\boxplus y \boxdot^{s+|x|-|y|}$.
\sp
It follows that the set 
$B = \{\conv(x,y): x \in dom(f) \wedge y = f(x)\}$
is regular as well, as it is first-order definable from
$A$ and the prefix relation: $z \in B$ $\Leftrightarrow$
$z$ does not end with $(\boxdot,\boxdot)$ and
$z \cdot (\boxdot,\boxdot)$ is a prefix of an element in $A$.
Thus $f$ is automatic.\niceqed

\begin{rem}
One might ask whether the condition on the input and output starting at
the same position is really needed. The answer is ``yes''. Assume by way
of contradiction that it would not be needed and that all functions linear
time computable by a one-tape Turing machine without any restrictions on
output positions are automatic. Then one could consider the free monoid
over $\{0,1\}$. For this monoid, the following function could be computed
from $\conv(x,y)$: The output is $z=f(x,y)$ if $y = xz$; the output is $\#$
if such a $z$ does not exist. For this, the machine just compares $x_1$ with
$y_1$ and erases $(x_1,y_1)$, $x_2$ with $y_2$ and erases $(x_2,y_2)$ and so on,
until it reaches (a) a pair of the form $(x_m,y_m)$ with $x_m \neq y_m$
or (b) a pair of the form $(x_m,\boxdot)$ or (c) a pair of the form
$(\boxdot,y_m)$
or (d) the end of the input.
In cases (a) and (b) the output has to be $\#$ and the machine just erases
all remaining input symbols and puts the special symbol $\#$ to denote the
special case; in case (c) the value $z$ is just obtained by changing
all remaining input symbols $(\#,y_k)$ to $y_k$ and the Turing machine
terminates. In case (d) the valid output is the empty string and the
Turing machine codes it adequately on the tape. Hence $f$ would be
automatic. But now one could first-order define concatenation $g$ by
letting $g(x,z)$ be the $y$ for which $f(x,y) = z$; this would give
that the concatenation is automatic, which is known to be 
false. The non-automaticity of the concatenation can be seen
as follows: For each automatic function there
is, by the pumping lemma \cite{HMU01}, a constant $c$ such that
each value is at most $c$ symbols longer than the corresponding input; now the
mapping $\conv(x,y) \mapsto xy$ fails to satisfy this for any given
constant $c$, for example, $x = 0^{c+1}$ and $y=1^{c+1}$ are
mapped to $xy = 0^{c+1}1^{c+1}$.
Hence the condition on the starting-positions cannot be dropped.
\end{rem}

\noindent
One can generalise non-deterministic computation to computation
by alternating Turing machines \cite{CKS81}.
Well known results in this field \cite{CKS81}
are that sets decidable in alternating logarithmic space are equal
to sets decidable in polynomial time and
that alternating polynomial time computations define the class
PSPACE for sets. Therefore it might be interesting
to ask what is the equivalent notion for alternating linear time computation.
The following definition deals with the
alternating computation counterpart of position-faithful linear
time computations.

\begin{adefn} \rm
An alternating position-faithful one-tape Turing machine $M$ has
$\exists$-states and $\forall$-states among the Turing machine 
states which permit the machine to guess one bit (which can then
be taken into account in future computation).
It uses, as the name says, exactly one tape which initially
contains $\boxplus x \boxdot^{\infty}$, where $x$ is the input string.
At the end of the computation, the output is the string 
between the $\boxplus$ and the first $\boxdot$.
$M$ is linear time bounded iff there is a constant $c$ such that, for
each input $x$ of length $n$ and each run of $M$, the duration of
the run until $M$ halts is at most $c\cdot (n+1)$ time steps.
Furthermore, $M$ alternatingly computes a function $f$
iff for each string $x$ on the input there is a unique string $y$ 
(which must be equal to $f(x)$) such that,
for a computation tree $T$ formed by chosing at each $\exists$-state
the guessed bit appropriately (the $\forall$-states are still
true branching nodes in this tree $T$), one has that
each computation path on $T$ ends up in an accepting state 
and each computation produces the same output $y$.
\end{adefn}

\noindent
It is easy to see that every function $f$ computed non-deterministically
by a position-faithful one-tape Turing machine in linear time
is also computed by an alternating position-faithful one-tape Turing
machine in linear time. However, the converse direction is open;
if the answer would be negative, one could use it as the basic definition
of a concept similar to automatic structures which is slightly more
general.

\begin{oprob}
Is every function $f$ computable in alternating linear time
by a position-faithful one-tape Turing machine automatic?
\end{oprob}

\section{Linear Time Learners} \label{se:lintimelearn}

\noindent
The following definition of learning is based on
the Gold's \cite{Go67} notion of learning in the limit. The presentation
differs slightly in order to incorporate memory restrictions and automaticity
as considered in this paper; note that learners without any restrictions
on the way the long term memory is organised can store all past data and
are therefore as powerful as those considered by Gold \cite{Go67}.

Informally, a learning scenario can be described as follows.
Suppose a family $\{L_e: e \in I\}$ of languages is given (in some effective
form), where $I$ is an index set.
The learner, as input, gets a listing of the elements of some set $L_e$.
The learner is supposed to figure out, 
in the limit from the listing as above, an index $d$ such that $L_d = L_e$. 
For ease of presentation it is assumed that all languages $L_e$
are not empty.
\sp
The listing of elements is formalised as a text.
A {\em text} $T$ for a language $L$ is an infinite sequence,
$w_0, w_1, w_2, \ldots$, containing all elements of $L$ but no
non-element of $L$, in any order with repetitions allowed. 
Let $T[n]$ denote the sequence of first $n$ elements of the text:
$w_0, w_1,\ldots, w_{n-1}$.
The basic model of inductive inference \cite{An80,BB75,Go67,JORS99,OSW86}
is that the learner $M$ is given a text $w_0,w_1,\ldots$
of all the words in a language $L$, one word per cycle.
At the same time $M$ outputs a sequence $e_0,e_1,\ldots$ of indices, 
one index in each cycle. Intuitively, each $e_i$ can be considered
as a conjecture of the learner regarding what the language $L$ is,
based on the data $w_0,w_1,\ldots,w_{i-1}$. In general, the indices
conjectured are from some index set $J$ and 
interpreted in a hypothesis space $\{H_e: e \in J\}$, where
$\{L_e: e \in I\} \subseteq \{H_e: e \in J\}$.
\sp
The learner maintains information about past data in
form of some memory, which may change between cycles.
Thus, the learner can be considered as an
\begin{quote}
algorithmic mapping from (old memory, new datum) to
(new memory, new conjecture)
\end{quote}
where the learner has some fixed initial memory.
The learner {\em learns} or {\em identifies} the language $L$, if,
for all possible texts for $L$, there is some $k$
such that $e_k$ is an index for $L$ and
$e_{k'}=e_k$ for $k' \geq k$. The learner learns a class $\CalL$ of
languages if it learns each language in $\CalL$.
\sp
The most basic set of hypothesis spaces are
automatic families of languages. Here, a family of
languages $\{L_e: e \in I\}$, is {\em automatic} if the index set $I$ and
the set $\{\conv(e,x): e \in I, x \in L_e\}$ are both regular. 
Automatic families \cite{JLS09,JOPS10} are the automata-theoretic 
counterpart of indexed families \cite{An80,LZZ08} which were widely used
in inductive inference to represent the class to be learnt.
Note that when $\{H_d: d \in J\}$ is a hypothesis space for $\{L_e: e \in I\}$,
which is an automatic family as well, then there is an automatic function $f$
mapping the indices from $J$ back to those in $I$, that is, $L_{f(d)} = H_d$
for all those $d \in J$ where $H_d$ equals some $L_e$. Hence one can
without loss of generality (for learning criteria considered in
this paper) directly use the hypothesis space
$\{L_e: e \in I\}$ for the class $\{L_e:e \in I\}$ to be learnt.
\sp
A learner $M$ is called {\em automatic}
if the mapping (old memory, new input word) to 
(new memory, new conjecture) for the learner is automatic, that is,
the set
$$
   \{\conv(om,dat,nm,nc): M(om,dat)=(nm,nc)\}
$$
is regular. In general, $om$ and $nm$ are the old and new versions
of the long term memory of the learner. Automatic learners are, roughly
speaking, the most restrictive form of learners which update a long term
memory in each cycle where they process one new datum given to the learner.
\sp
The next definition generalises the notion of automatic learning to
a learner which has a linear or nearly linear time bound for each of
its cycle. This generalisation is natural, due to the correspondence
between automatic function and linear time computable functions given
in the previous section of this paper.

\begin{adefn}
A learner $M$ is a Turing machine which maintains some memory and in each
cycle receives as input one word to be learnt, updates its memory
and then outputs an hypothesis.
The tapes of the Turing machine are all one-sided
infinite and contain $\boxplus$
at the left end. The machine operates in cycles, where in each cycle
it reads one current datum (from a text of the language to be learnt)
and formulates one hypothesis. Furthermore, it has some long term memory
in its tape where the memory in Tape $0$ is always there while the
memories in the additional data structures (Tapes $1,2,\ldots,k$)
is only there when these additional data structures are explicitly
permitted.
\begin{iteMize}{$\bullet$}
\item At the beginning of each cycle,
      Tape $0$ (base tape) contains convolution of the input
      and some information (previous long term memory) which is not 
      longer in length (up to an additive
      constant) than the length of the longest word seen so far.
      The head on Tape $0$ of the Turing machine starts at $\boxplus$ at
      the beginning of each cycle.
\item At the end of each cycle,
      Tape $0$ (base tape) has to contain the convolution of
      the new long term memory and the hypothesis which the
      learner is conjecturing.
\item During the execution of a cycle, the learner
      can run in time linear in the current length of Tape $0$
      and, like a position-faithful one-tape Turing machine,
      replace the convolution of the current datum and old long term
      memory by the convolution of the hypothesis and the new long term
      memory. Furthermore, the memory in Tape $0$ 
      has to meet the constraint that
      it is at most (up to an additive constant) the length of the longest 
      datum seen so far (including the current datum), hence there
      is an explicit bound on the length of Tape $0$ in each cycle.
\item Tapes $1,2,\ldots,k$ are normal tapes, whose contents and head positions
      are not modified during change of cycles. $M$ can use these tapes
      for archiving information and doing calculations. There is
      no extra time allowance for the machine to use these tapes, hence
      the machine can only access a small amount (linear in the length
      of Tape $0$) in each cycle of these tapes.
\item Without loss of generality, one can assume that the length of the
      longest datum seen so far is stored in the memory in Tape $0$.
\end{iteMize}
The learner is said to have $k$ additional work tapes iff
it has in addition to Tape $0$ also the Tapes $1,2,\ldots,k$.
Figure~\ref{figdefnlearner} illustrates a learner with two additional tapes.
\end{adefn}

\begin{figure}[t]
\begin{tikzpicture}[box/.style={draw,rectangle,minimum width=45pt,
  minimum height=45pt},
  start chain=1 going right,
  start chain=2 going right,
  start chain=3 going right,
  node distance= 0.35mm]
    \node [on chain=1] {Base Tape \ \mbox{ }};
    \node [on chain=2, yshift=-1cm] {Work Tape 1};
    \node [on chain=3, yshift=-2cm] {Work Tape 2};
    \node[draw,on chain=1]{\mbox{$\boxplus$}};
    \foreach \x in {1,2} {
        \x, \node [draw,on chain=1] {$C$}; }
    \node[draw,on chain=1]{$H$};
    \foreach \x in {1,2} {
        \x, \node [draw,on chain=1] {$C$}; }
    \node[draw,on chain=1]{\mbox{$\boxdot$}};
    \node[draw,on chain=2]{\mbox{$\boxplus$}};
    \foreach \x in {1,2,...,3} {
        \x, \node [draw,on chain=2] {$C$}; }
    \node[draw,on chain=2]{$H$};
    \foreach \x in {1,2,...,5} {
        \x, \node [draw,on chain=2] {$C$}; }
    \foreach \x in {1,2,...,3} {
        \x, \node [draw,on chain=2] {$\boxdot$}; }
    \node[draw,on chain=3]{\mbox{$\boxplus$}};
    \foreach \x in {1,2,...,5} {
        \x, \node [draw,on chain=3] {$C$}; }
    \node[draw,on chain=3]{$H$};
    \foreach \x in {1,2,...,4} {
        \x, \node [draw,on chain=3] {$C$}; }
    \foreach \x in {1,2} {
        \x, \node [draw,on chain=3] {$\boxdot$}; }
    \node [name=r,on chain=2] {\ldots}; 
    \node [name=r,on chain=3] {\ldots}; 
\end{tikzpicture}
\caption{Learner with two working tapes; the head positions are $H$ and other
   data positions are $C$; note that data characters can be convoluted
   characters from finitely many alphabets in order to store the convolution
   of several tracks on a tape.} \label{figdefnlearner}
\end{figure}

\noindent
Note that in the definition of Tape $0$, it is explicit that the length of
the hypothesis produced is bounded by the length of the largest
example seen so far plus a constant. This is compatible with learning, as 
for all automatic families, (i) for any finite set $L$ in the family,
the length of the smallest index $e$ for $L$ overshoots the length 
of the longest element of $L$ by at most a constant
and (ii) for any infinite set $L$ in the family
there are words in $L$ which are longer than some index for $L$;
thus a learner cannot fail just because
the indices of a language $L_e$ are extremely long compared
to the size of the members of $L_e$ -- though, of course, there may be other
reasons for a learner not to be successful.
\sp
Note that if
only Tape $0$ is present, the model is equivalent to an automatic learner
with the memory bounded by the size of the longest datum seen so
far (plus a constant) \cite{CJLOSS11,JLS09}. The next examples illustrate
what type of learnable automatic classes exist.

\begin{exa} \label{ex:autolearn}
The following automatic classes are learnable by an automatic learner
with its memory bounded by the length of the longest example seen so
far (plus a constant):
\sp
First, the class of all extensions of an index, that is,
$I = \Sigma^*$ and $L_e = e \cdot \Sigma^*$
for all $e \in I$. Here the learner maintains as a memory the longest
common prefix $e$ of all data seen so far and whenever the memory is $e$
and a new datum $x$ is processed, the learner updates $e$ to the longest
common prefix of both, $e$ and $x$, which is also the next hypothesis.
\sp
Second, the class of all closed intervals in the
lexicographic ordering, that is, $I = \{conv(d,e): d,e \in \Sigma^*
\wedge d \leq_{lex} e\}$ and $L_{conv(d,e)} = \{x \in \Sigma^*:
d \leq_{lex} x \leq_{lex} e\}$; here $x \leq_{lex} y$ denotes that
 $x$ is lexicographically before $y$.
The learner maintains as memory
the lexicographically least and greatest elements seen so far, the convolution
of these elements also serves as hypothesis.
\sp
Third, the class of all strings of
length different from the index, that is, $I = \{0\}^*$ and
$L_e = \{x \in \Sigma^*: |x| \neq |e|\}$. Here the learner archives in
Tape $0$ binary string which is of length $1$ plus the length of
the longest example
seen so far; the $k$-th bit of this string (starting with $k=0$)
is $1$ iff an example
of length $k$ has been seen so far, and $0$ iff no example
of length $k$ has been seen so far. The conjecture is $0^h$ for the least
$h$ such that either the $h$-th bit of the memory is $0$ or $h$ is $1$ plus
the length of the memory string.
\end{exa}

\noindent
For any automatic family, $\{H_e: e \in J\}$, 
the equivalence question for indices is automatic, that is,
the set $\{\conv(e,e'): H_e=H_{e'}\}$ is regular.
Thus for the purposes of this paper, 
one can take the hypothesis space to be one-one, 
that is, different indices represent different languages.
In a one-one hypothesis space, the index $e$ of a finite language $L_e$
has, up to an additive constant, the same length as the longest word in $L_e$;
this follows easily from \cite[Theorem 3.5]{JOPS10}.
This observation is crucial as otherwise the time-constraint on the learner
would prevent the learner from eventually outputting the correct index;
for infinite languages this is not a problem as the language must contain
arbitrarily long words.
\sp
Angluin \cite{An80} gave a characterisation when a class is learnable in 
general. This characterisation, 
adjusted to automatic families, says that a class is learnable
iff, for every $e \in I$, there exists a finite set $D \subseteq L_e$
such that there is no $d \in I$ with $D \subseteq L_d \subset L_e$. All
the automatic families from Example~\ref{ex:autolearn} satisfy this criterion;
however, Gold \cite{Go67} provided a simple example of a non-learnable
class which of course then also fails at Angluin's criterion: One infinite
set plus all of its finite subsets.
\sp
The main question of this
section is which learnable classes can also be learnt by a linear-time learner
with $k$ additional work tapes. For $k=0$,
this is in general not possible, as 
automatic learners fail to learn various learnable classes \cite{JLS09},
for example the class of all sets $\{0,1\}^*-\{x\}$, with the index
$x$ being from $\{0,1\}^*$, and the class of all
sets $L_e = \{x \in \{0,1\}^{|e|}: x \neq e\}$.
\sp
Freivalds, Kinber and Smith \cite{FKS95} introduced limitations on the
long term memory into inductive inference; Kinber and Stephan \cite{KS95}
transferred it to the field of language learning. Automatic learners have
similar limitations and are therefore not able to learn all learnable
automatic classes \cite{CJLOSS11,JLS09}.
The usage of additional work tapes for linear time learners permits to
overcome these limitations, the next results specify how many additional
work tapes are needed. 
Recall from above that work tapes are said to be {\em additional\/}
iff they are in addition to
the base tape.

\begin{thm} \label{th:onetape}
Suppose $\Sigma=\{0,1,2\}$ and consider the automatic family
$\CalL$ over the alphabet $\Sigma$ which is defined as follows:
$\CalL$ consists of
(i) $L_{\varepsilon} = \{0,1\}^*$ and
(ii) $L_{x0} = \{0,1\}^*\cup \{x2\}-\{x\}$ and 
(iii) $L_{x1}=\{0,1\}^*\cup\{x2\}$,
for each $x \in \{0,1\}^*$.
Then, $\CalL$
does not have an automatic learner but has a linear-time learner using one
additional work tape.
\end{thm}

\proof
An automatic learner cannot memorise all the data from $\{0,1\}^*$ it sees.
For any automatic learner, one can show, see \cite{JLS09}, that there
are two finite sequences
of words from $L_{\varepsilon}$, one containing $x$ and one not
containing $x$, such that the
automatic learner has the same long term memory after having seen both
sequences.
If one presents to the automatic learner, after these sequences, all
the elements
of $L_{x0}$, then the automatic learner's limiting behaviour on the
two texts so formed is the same, even though they are texts for two different
languages, $L_{x1}$ or $L_{x0}$, in $\CalL$.
Therefore the learner cannot learn the class $\CalL$.
\sp
A linear time learner with one additional work tape (called Tape $1$)
initially conjectures $L_{\varepsilon}$ and uses Tape $1$ to archive 
all the examples seen at the current end of the written part of the tape.
When the learner sees a word of the form $x2$, it maintains a copy of
it in the memory part of Tape $0$ and conjectures $x0$ as its hypothesis. 
In each subsequent cycle, the
learner scrolls back Tape $1$ by one word and compares the word there
as well as the current input with $x2$; if one of these two is $x$ then
the learner changes its conjecture to $L_{x1}$, else it keeps its
conjecture as $L_{x0}$.
In the case that the origin of Tape 1 ($\boxplus$) is reached, the learner from
then onwards ignores Tape 1 and only compares the incoming input
with $x2$.
It is easy to verify that the learner as described above learns~$\CalL$.\niceqed

\begin{thm} \label{th:twotapes}
Every learnable automatic family $\CalL$ has a linear-time learner using
two additional work tapes.
\end{thm}

\proof
Jain, Luo and Stephan \cite{JLS09} showed that for every learnable
automatic family $\CalL=\{L_e: e \in I\}$ there is an automatic
learner $M$ using memory bounded in length by the length 
of the longest example seen
so far (plus a constant) which learns the class from every fat text
(a text in which every element of the language appears infinitely often). 
So the main idea is to use the two additional tapes in order to 
simulate and feed the learner $M$ with a fat text. 
The two additional tapes are used to store all the incoming data and
then to feed the learner $M$ with each data item infinitely often.
The words in the tapes are stored using some separator $\#$ to separate
the words. Thus, $00\#1\#\#11\boxdot$ indicates that the tape
contains the words $00$, $1$, $\varepsilon$ and $11$.

The learner $N$ for $\CalL$ using two additional tapes works as follows.
Suppose the previous memory stored in Tape 0 is $mem_k$ (initially the memory
stored on Tape $0$ is the initial memory of $M$)
and the current datum is $w_k$. Then, $N$ does the following:
\begin{iteMize}{$\bullet$}
\item Compute $M(mem_k,w_k)=(mem',e')$.
\item Find the last word in Tape $1$, say $t$. Erase this word from Tape $1$.
      In the case that Tape $1$ was already empty, let $t=w_k$.
\item Compute $M(mem',t)=(mem_{k+1},e_k)$.
\item Write $w_k$ and $t$ at the end of Tape $2$ (using the separator $\#$ to
      separate the words).
\item When the beginning of Tape $1$ is reached ($\boxplus$), interchange 
      the roles of Tape $1$ and Tape $2$ from the next cycle.
\item The new memory to be stored on Tape $0$ is $mem_{k+1}$ and 
      the conjecture is $e_k$.
\end{iteMize}
It is easy to see that in each cycle, the time spent is
proportional to $|mem_k|+|w_k|+|t|$ and thus linear in the length
of the longest word seen so far (plus a constant); note that 
$mem',e',e_k$ are also bounded
by that length (plus a constant). Furthermore, in the simulation of
$M$, each input word to $N$ is given to $M$ infinitely often.
Hence $N$ learns each language from the class $\CalL$.\niceqed

\begin{oprob}
It is unknown whether one can learn every in principal learnable
automatic class using an automatic learner augmented by only one
work tape.
\end{oprob}

\medskip
\noindent
Further investigations deal with the question what happens if one does
not add further work tapes to the learner but uses other methods to
store memory.
Indeed, the organisation in a tape is a bit awkward and using a queue solves
some problems. A queue is a tape where one reads at one end and writes at the
opposite end, both the reading and writing heads are unidirectional and cannot
overtake each other. Tape $0$ satisfies the same constraints as in the model
of additional work tapes and one also has the constraint that in each cycle
only linearly many symbols (measured in the length of the longest datum seen
so far) are stored in the queue and retrieved from it.

\begin{thm} \label{th:queue}
Every learnable automatic family $\CalL$ has a linear-time learner using
one additional queue as a data structure.
\end{thm}

\proof
The learner simulates an automatic learner $M$ for $\CalL$
using fat text, in a way similar
to that done in Theorem~\ref{th:twotapes}. Let $M$ in the $k$-th step map
$(mem_k,w_k)$ to $(mem_{k+1},e_k)$ for $M$'s memory $mem_k$. 
\sp
For ease of presentation, the contents of Tape $0$ is considered as
consisting of a convolution of 4 items (rather than $2$ items,
as considered in other parts of the paper).
At the beginning of a cycle the linear-time learner $N$ has 
$\conv(v_k,-,mem_k,-)$ on Tape $0$
where $v_k$ is the current datum, $mem_k$ the archived memory of $M$ and
``$-$'' refers to irrelevant or empty content. In the $k$-th cycle,
the linear-time learner $N$ 
scans four times over Tape $0$ from beginning to the end
and each time afterwards returns to the beginning of the tape:
\begin{enumerate}[(1)]
\item Copy $v_k$ from Tape $0$ to the write-end of the queue;
\item Read a word from the read-end of the queue, call it $w_k$,
      and update Tape $0$ to $\conv(v_k,w_k,\linebreak[3] mem_k,-)$;
\item Copy $w_k$ from Tape $0$ to the write-end of the queue;
\item Simulate $M$ on Tape $0$ in order to map $(mem_k,w_k)$ to
      $(mem_{k+1},e_k)$ and update Tape $0$ to $\conv(v_k,w_k,mem_{k+1},e_k)$.
\end{enumerate}
It can easily be verified that this algorithm permits to simulate $M$
using the data type of a queue and that each cycle takes only time
linear in the length of the longest datum seen so far.
Thus, $N$ learns $\CalL$. \niceqed

\medskip
\noindent
A further data structure investigated is the provision of
additional stacks. Tape $0$ remains a tape in this model and has
still to obey to the resource-bound of not being longer than the longest
word seen so far (plus a constant). Theorems~\ref{th:onetape}
and~\ref{th:twotapes} work
also with one and two stacks, respectively, as the additional work tapes
are actually used like stacks.

\begin{thm}
There is an automatic class which can be learnt
with one additional stack but not by an automatic learner. Furthermore,
every learnable automatic class can be learnt by a learner
using two additional stacks.
\end{thm}

\noindent
Furthermore, the next result shows that in general one stack is not enough;
so one additional stack gives only intermediate learning power while two
or more additional stacks give the full learning power. The class witnessing
the separation contains only finite sets.
\sp
For information on Kolmogorov complexity,
the reader is referred to standard text books \cite{Ca02,DH10,LV08,Ni09}. 
The next paragraphs provide a brief description of the basic concepts.
\sp
Consider a Turing machine $U$ which computes a partial-recursive function from 
$\{0,1\}^* \times
\{0,1\}^*$ to $\{0,1\}^*$. The first input to $U$ 
is also referred to as a program.
Machine $U$ is universal iff for every further
machine $V$, there is a constant $c$ such that, for every $(p,y)$
in the domain of $V$, there is a $q$, which is at most $c$ symbols longer than
$p$, satisfying $U(q,y)= V(p,y)$. 
Fix a universal machine $U$. Now the conditional
Kolmogorov complexity $C(x|y)$ is the length of the shortest program
$p$ with $U(p,y) = x$; the plain Kolmogorov complexity $C(x)$
is $C(x|\varepsilon)$. Note that, due to the universality of $U$,
the values of $C(\cdot)$ can only be improved by a constant 
(independent of $x,y$)
when changing from one universal machine to another one.
In some cases below, $C(x| y_1,y_2,\ldots,y_r)$ is the conditional
Kolmogorov complexity when given an $r$-tuple $(y_1,y_2,\ldots,y_r)$
where $r$ might vary; such a tuple can be coded up in any way which
permits to identify the parts uniquely, as automaticity is not required,
the coding $0^{|y_1|} 1 y_1 0^{|y_2|} 1 y_2 \ldots 0^{|y_r|} 1 y_r$ would do it.
\sp
If $f$ is a partial-recursive function then there is a constant $c$ with
$C(f(x)|y) \leq C(x|y)+c$ for all $x$ in the domain of $f$. In particular,
if one can find a way to describe the strings $x$ in a set $A$
by binary strings $p_x$
such that some algorithm can compute each $x \in A$
from the corresponding description $p_x$, then $C(x) \leq |p_x|+c$
for some constant $c$ and all $x \in A$. For this reason, one often
says that $x$ can be described by $n$ bits when the corresponding
$p_x$ can be chosen to have $n$ bits.

\begin{thm} \label{th:stackminusage}
The class of all $L_e = \{x \in \{0,1\}^{|e|}: x \neq e\}$ with
$e \in \{0,1\}^* \cup \{2\}^*$ cannot be learnt by a linear-time learner using
one additional stack.
\end{thm}

\proof
Assume that the linear-time learner $M$ using one stack, 
in addition to the base tape,
is given.  In order to find languages not
learnt by $M$, one focuses on $L_e$ where the parameter $n = |e|$ is
large; in addition one considers only $n$ of the form $k+2^k$ for some $k$;
this parameter $k$ and $m = 2^k$ will play some role in the arguments below.
Note that all data-items in $L_e$ have the length $n$.
For $i \in \{1,2,\ldots,m\}$, let $x_i$ be a
string of length $n$ such that the Kolmogorov
complexity $C(x_1 x_2 \ldots x_m)$ is at least $(n-k)m$ and the first $k$ bits
of each $x_i$ is the binary bit representation of $i-1$. Furthermore,
assume that $c$ is a constant so large that the Kolmogorov complexity
of the content of Tape $0$ (which can be assumed to be always $n$
symbols long, since
all data have length $n$, but which can use more than two alphabet symbols)
is at most $cn$ and that the stack can, in each round, pull or push
up to $cn$ symbols, where the stack alphabet has at most $2$ symbols
(one can code up a larger alphabet in binary and choose the constant
$c$ sufficiently large to absorb the extra amount of storage).
Hence, in each cycle, what the machine does depends on the content of
Tape $0$ (worth $cn$ bits) and on the top $cn$ symbols of the stack
(worth $cn$ bits). Furthermore, assume that all words of length $n$
different from $x_1,x_2,\ldots,x_m$ have already been presented to the learner
and let $\alpha$ denote the content of Tape $0$ and $u\beta$ denote the content
of the stack where $\beta$ are the top $cn$ symbols (or less if $u$ is the
empty word). Below one considers the behaviour/configuration of the learner
when it is presented with further inputs and one considers $(\alpha,u\beta)$
as the initial configuration of the learner for this purpose.
Below, the configuration of the learner, at any stage before reading the
next input, is denoted by $(\cdot,\cdot)$, where the first argument
is the content of the tape and the second argument is the content
of the stack.
\sp
Intuitively, as the $x_j$'s are complex, the learner needs to store
them on the stack when it receives them (otherwise, it would lose
information about which $x_j$'s it has seen). This forces the stack
to grow larger and larger and prevents the learner from accessing
earlier stored data on the stack, thus making the earlier stored
information useless. This allows to show that some language $L_e$
is not learnable by $M$.
\sp
Claim~\ref{cl:third} gives a permutation 
$x_{i_1},x_{i_2},\ldots,x_{i_m}$ of $x_1,x_2,\ldots,x_m$ such that
$M$ does not touch any, but the top $6(c+1)^2n$ symbols of $u$, on
input $x_{i_1},x_{i_2},\ldots,x_{i_m}$. Claim~\ref{cl:fourth} uses
this claim to show that on some sequence $\sigma$ of $x_i$'s, $M$ reaches a
configuration $(\alpha',vw'\beta')$, with $|\beta'| =cn$ and
$v$ being same as $u$ except for the top $6(c+1)^2n$ symbols removed,
where the learner never touches $v$ on the input 
$\sigma$, and for any future input involving $x_i$'s never
touches $vw'$. This, then allows to claim in Claim~\ref{cl:fifth}
that the learner cannot learn some $L_e$. 
Claims~\ref{cl:first} and~\ref{cl:second} are used in proving the above
claims.
Now the five claims about the configuration of $M$ are proven formally.

\begin{clm} \label{cl:first}
There do not exist two distinct input sequences of words of length
$n$, one containing an $x_i$ 
and one not containing $x_i$, ending up in the same configuration
$(\alpha',v\beta')$ where $\beta'$ has at most length $cn$ and $v$ has not been
touched (that is, starting from configuration $(\alpha,u\beta)$, the
initial portion 
$v$ of $u\beta$ above 
was never at the top of the stack during the processing of any of the
two sequences).
\end{clm}

\noindent
Assume by way of contradiction that this claim fails, that is, 
there are two such sequences $\sigma$ and $\sigma'$. Then one can
bring the learner into the
configuration $(\alpha',v\beta')$ by either of the sequences and thereafter feed
the learner with the $x_j$ with $j \neq i$, and then with a string 
of length $n$, different from $x_i$, forever.
The convergence
behaviour of the learner, in both cases, is the same as the configuration
$(\alpha',v\beta')$ is independent of the sequence $\sigma$ or $\sigma'$
by which the learner reached it;
from then onwards the learner receives, in both cases, the same data and 
conjectures the same hypotheses, as in both cases they are based on 
the same data, Tape 0 and stack.
In one case the learner has to learn $L_{x_i} = \{0,1\}^n-\{x_i\}$ and in
the other case the learner has to learn $L_{2^n} = \{0,1\}^n$; thus
the learner can learn at most one of these two sets.
This completes the proof of the claim.

\begin{clm} \label{cl:second}
There is no input sequence $(x_{i_1},x_{i_2},\ldots,x_{i_\ell})$
and no splitting of $u$ into $vw$ such that $M$, after reading these
inputs, is in a configuration of the form $(\alpha',v\beta')$ with
$|\beta'| \leq cn$ and without having pulled and pushed back any
symbols of $v$ and with the conditional Kolmogorov complexity satisfying
$C(x_{i_1} \linebreak[3] x_{i_2} \linebreak[3] \ldots \linebreak[3]
x_{i_\ell}| \alpha ,w\beta, \linebreak[3] i_1,i_2,\ldots,i_\ell) 
\linebreak[3] \geq
(c+1)^2n$.
\end{clm}

\noindent
For a proof of the claim,
assume by way of contradiction that there is such an input sequence 
$(x_{i_1},x_{i_2},\ldots,x_{i_{\ell}})$.
Then there is a partial-recursive
function $f$ such that $f$, given 
$(\alpha,w\beta,i_1,i_2,\ldots,\linebreak[3]
i_\ell,\alpha',\beta')$, finds a sequence
$y_{i_1},y_{i_2},\ldots,y_{i_\ell}$ such that $y_{i_j} = y_{i_{j'}}$ whenever
$i_j = i_{j'}$, $y_{i_j} \in \{0,1\}^n$ for all $j$, $y_{i_j}$ having
the first $k$ bits being the binary representation of $i_j$ and
$M$ pulling on these inputs the symbols belonging to $w\beta$ without
touching those of $v$ and ending up in the configuration $(\alpha',v\beta')$.
Note that one does not need to know $v$ for this search, hence the
search depends only on the inputs given to $f$ and returns an input
sequence such that its Kolmogorov complexity 
given
$(\alpha,w\beta,i_1,i_2,\ldots,i_\ell)$ is at 
most that of $(\alpha',\beta')$, that
is, below $(c+1)^2n$ (assuming that $n$ is sufficiently large).
It follows that at least one $y_{i_j}$ differs from $x_{i_j}$;
furthermore, no other $y_{i_{j'}}$ can be equal to $x_{i_j}$
by the rules that each $y_{i_{j'}}$ encodes $i_{j'}$ in the
first $k$ bits and equals to $y_{i_j}$ whenever $i_{j'} = i_j$.
However, this would contradict Claim~\ref{cl:first}.
This completes the proof of
Claim~\ref{cl:second}.

\begin{clm} \label{cl:third}
There is a permutation $(x_{i_1},x_{i_2},\ldots,x_{i_m})$ of
$(x_1,x_2,\ldots,x_m)$ such that the splitting $vw = u$ with
either $|w|=6(c+1)^3n$ or $|w|<6(c+1)^3n \wedge |v|=0$
satisfies that $M$ on input $(x_{i_1},x_{i_2},\ldots,x_{i_m})$
never touches the symbols in $v$.
\end{clm}

\noindent
Let $vw$ be the given splitting of $u$.
If $|w| < 6(c+1)^3n$ then $v$ is empty and nothing needs to be proven;
thus assume that $|w| = 6(c+1)^3n$.
\sp
Now define $y_k = x_1 x_2 \ldots x_m$
and inductively for $\ell = k-1,k-2,\ldots,1$, split $y_{\ell+1}$ at the
middle into two equal parts $y_{\ell}$ and $z_\ell$ 
$($both of length $2^{\ell}n)$ such that
$C(y_\ell | k,\alpha,w\beta) \geq C(z_\ell | k,\alpha,w\beta)$. Note
that there is a unique permutation
of the form $(x_{i_1},x_{i_2},\ldots,x_{i_m})$ of
$(x_1, \linebreak[3] x_2, \linebreak[3] \ldots, \linebreak[3] x_m)$
such that
$$
   x_{i_1} x_{i_2} \ldots x_{i_m} = y_1 z_1 z_2 \ldots z_{k-1}.
$$
Note that $i_2,i_3,\ldots,i_m$ can be computed from $i_1$.
Note that $C(y_k | k,\alpha,w\beta) \geq (n-2k)m$ 
(for $k$ and $n = k+2^k$, $m=2^k$ being sufficiently large)
for the following reasons:
$C(y_k) \geq (n-k)m$;
$C(y_k | k,\alpha,w\beta) \geq C(y_k)-C((k,\alpha,w\beta))-k$;
$C(k,\alpha,w\beta)+k \leq 8(c+1)^3n \leq km/2$.
\sp
By induction one can see that
$C(y_\ell | k,\alpha,w\beta) \geq (n-2k)m \cdot 2^{\ell-k}-k$ for all $\ell$
whenever $k,n$ are sufficiently large; note that $y_{\ell}$ is the more
complex half of $y_{\ell+1}$ and therefore has by induction hypothesis
at least the complexity $(n-2k)\linebreak[3] \cdot m \cdot 2^{\ell+1-k}/2-k/2$
minus some constant
which can be brought into the form $(n-2k)m \cdot 2^{\ell-k}-k$ by
assuming that $k/2$ is larger than the corresponding constant.
\sp
Furthermore, the values $i_1,i_2,\ldots,i_m$ can be computed from $k$ and
$i_1$, hence one can represent $i_1,i_2,\ldots,i_h$
by $i_1$ and $h$ and $k$. Hence
$$
   C(y_\ell | \alpha,w\beta,i_1,\ldots,i_h) \geq (n-2k)m \cdot 2^{\ell-k} - 5k
$$
for any $h$ with $2^{\ell} \leq h < 2^{\ell+1}$.
There are two cases for each $h$ with $2^{\ell} \leq h < 2^{\ell+1}$:
\sp
First, $3(c+1)^2 n > |y_\ell|$. Then $h < 6(c+1)^2$ and,
on input $(x_{i_1},x_{i_2},\ldots,x_{i_h})$, the learner
can have pulled at most $6c(c+1)^2n$ symbols
from the stack; hence it has neither touched $v$ nor the bottom
$cn$ symbols of $w$.
\sp
Second, $3(c+1)^2 n \leq |y_\ell|$. Then 
\begin{eqnarray*}
   C(y_\ell | \alpha,w\beta,k,i_1,i_2,\ldots,i_h) & \geq &
   (n-2k)m \cdot 2^{\ell-k} - 5k \ \geq \\
       3(c+1)^2(n-2k) - 5k & > & 2(c+1)^2n.
\end{eqnarray*}
Assuming that $k$ and $n = 2^k+k$ are sufficiently large, one obtains
$$
   C(x_{i_1}x_{i_2}\ldots x_{i_h}| \alpha,w\beta,i_1,i_2,\ldots,i_h)
\geq (c+1)^2n.
$$
Thus, using Claim~\ref{cl:second} it follows that for all 
$h \in \{6(c+1)^2,6(c+1)^2+1,\linebreak[3] \ldots,m\}$ 
there are at least $cn$ symbols in the stack above $v$ after reading
$x_{i_1},x_{i_2},\ldots,x_{i_h}$.
\sp
Hence, using above cases, one can conclude by induction on $h$ that the
symbols in $v$ are not touched while processing the input
$(x_{i_1},x_{i_2},\ldots,x_{i_m})$.

\begin{clm} \label{cl:fourth}
Split $u$ into $vw$ as in Claim~\ref{cl:third}. There is a sequence
of all $x_i$, perhaps with repetitions, such that after reading this
sequence $M$ is in a configuration $(\alpha',vw'\beta')$, with $|\beta'|=cn$, 
such that for all further
inputs from $x_1,x_2,\ldots,x_m$, $M$ does not touch the symbols on the
part of the stack denoted by $vw'$.
\end{clm}

\noindent
Assuming that this sequence does not exist, one could use the sequence
given in Claim~\ref{cl:third} to remain above $v$ in the stack until all
symbols are passed and then one could feed some sequence of $x_i$ until
all but at most $cn$ symbols above $v$ are used up; that is, one would
be in a configuration of the form $(\alpha',v\beta')$ with $|\alpha'| = n$ and
$|\beta'| \leq cn$. Hence one can, given $(\alpha,w\beta)$ and
$(\alpha',\beta')$ search a
tuple $(y_1,y_2,\ldots,y_m)$ such that each $y_i$ starts with a binary
number of length $k$ representing $i-1$ and each $y_i$ has $n$ bits
and there is a sequence of inputs drawn from this tuple on which the
configuration of $M$ with $(\alpha,vw\beta)$ changes to
$(\alpha',v\beta')$ without
touching $v$. The first tuple $(y_1,y_2,\ldots,y_m)$ of this type
found by searching has Kolmogorov complexity at most
$8(c+1)^3n$ (obtained by coding the inputs $k$, $\alpha$, $w\beta$,
$\alpha'$, $\beta'$ and
the routine for the search programme) which is less than 
$(n-k)m$, the lower bound on the 
Kolmogorov complexity of $x_1 x_2 \ldots x_m$, for sufficiently large $k,m,n$.
Therefore some $y_i$ differs from $x_i$ and therefore one can reach
the configuration $(\alpha',v\beta')$ from $(\alpha,vw\beta)$ by
either having seen
$x_i$ or not having seen $x_i$. It follows from Claim~\ref{cl:first}
that this cannot occur, hence there is some minimal extension $w'\beta'$
of $v$ such that $|\beta'| = cn$ and when reading any sequence of the
data $x_1,x_2,\ldots,x_m$ after having reached the
configuration $(\alpha',vw'\beta')$, it will not touch $vw'$ in the stack,
that is, all future activity depends only on $\alpha'$ and $\beta'$.

\begin{clm} \label{cl:fifth}
$M$ fails to learn some language of the form
$\{0,1\}^n-\{x_i\}$ or $\{0,1\}^n$.
\end{clm}

\noindent
Let $u,v,w,\alpha,\beta,\alpha',\beta',w'$ as in
Claim~\ref{cl:fourth}. One can now
show that there is a tuple $(y_1,y_2,\ldots,y_m)$ with
$|y_i| = n$ and $y_i$ extending the $k$-bit representation of $i-1$
such that $M$ when fed with some input-sequence taken from
$\{y_1,y_2,\ldots,y_m\}$ ends up in a
configuration of the form $(\alpha',vw''\beta')$ without touching $v$ and
this configuration is computed from $(\alpha,w\beta,\alpha',\beta')$; as in
Claim~\ref{cl:fourth} one can argue that some $y_i \neq x_i$.
Now one can feed all the $x_j \neq x_i$ into $M$ for the
configurations $(\alpha',vw'\beta')$ and $(\alpha',vw''\beta')$, respectively,
for both in the same way and in a loop repeated forever.
In both cases the learner $M$
either converges to the same index or does not converge,
but in one case the text which $M$ has received is a text for $\{0,1\}^n$
and in the other case it is a text for $\{0,1\}^n-\{x_i\}$.
Hence $M$ fails to learn at least one of these two sets.\niceqed

\section{Relaxing the Timing Constraints}

\noindent
In this section, it is investigated how the learning power improves
if the severe restrictions on work Tape $0$ or the computation
time are a bit relaxed.
The next result shows that, if one allows a bit more than just linear time,
then one can learn, using one work tape, all learnable
automatic classes of infinite languages. 
The result could even be transferred to families of arbitrary
r.e.\ sets as the simulated learner is an arbitrary recursive learner.
Intuitively, think of $f$ in the following theorem as a slowly growing
function.

\begin{thm} \label{th:onetapesuperlinear}
Assume that $\{L_e: e \in I\}$ is an automatic family where every $L_e$
is infinite and $M$ is a recursive learner which learns this family.
Furthermore, assume that $f,g$ are recursive functions with the property
that $f(n) \geq m$ whenever $n \geq g(m)$ $($so $g$ is some type of inverse
of $f)$. Then there is a learner $N$ which learns the above family,
using only one additional work tape,
and satisfies the following constraint: 
if $n$ is the length of the longest example seen so
far, then only the cells number $1,2,\ldots,n$ of Tape $0$ can be non-empty and
the update time of $N$ in the current cycle is $O(n \cdot f(n))$.
\end{thm}

\proof
The main idea of the proof is that one constructs a learner which splits
Tape $1$ into four 
tracks for archivation; the learner 
usually uses Track $1$; in irregular intervals, the learner returns from
its current position to the origin of Tape $1$ and uses Track $2$
for archiving the examples which come up during this ``return to
the origin'' until it reaches the old data on Tracks $2$ and $3$.
When this happens, the old data found there consist only of words
up to length $m$ (where $m$ is sufficiently small compared to the
current word length $n$) and the learner can compress the data in
Tracks $1$, $2$ and $3$ into a list $\alpha$ (to be maintained on
Tape $0$); $\alpha$ will contain, for each word $w$ up to length $m$ occurring
in the input, at most one copy (which gives a corresponding length bound
on the length of $\alpha$). Once the compression is completed, the
learner returns to the forward mode using the one left over free
track for this purpose. The key idea is to ``space out'' the visits
to the origin such that, for $m$ being the length of the longest datum
seen up to the end of the last visit, $m$ is so much smaller
than the current $n$ that $2^{m+1} \cdot (m+1) \leq f(n)$; this allows
all the data which was archived up to the end
of the previous visit to be compressed into a string of length
up to $f(n)$ and the update of this compressed memory can, in each
round, be done in time $O(f(n) \cdot n)$.
\sp
The description below gives a more precise description of the
update protocol. As the memory has only to be bounded by the length
of the longest datum seen so far plus some constant, one can assume
without loss of generality that $n$ is at least $1$.
\sp
On Tape $0$, as memory, the learner $N$ archives the convolution of variables
$\alpha,\beta,\gamma,e,\linebreak[3] 0^m,0^n$ with the following meaning.
\begin{iteMize}{$\bullet$}
\item $0^n$ represents in unary the length of the longest word seen so far
and $0^m$ is an old value of $0^n$; initially $m$ and $n$ are $1$ (not $0$).
\item The variable $\alpha$ is, during the runtime,
only modified by appending symbols at the end and will in the limit
consist of a one-one text of all the words occurring in the language
to be learnt; the words on $\alpha$ are separated by a special character.
For example, $\alpha = \# 0 0 \# \# 0 1 0 1 \# 1 1 1 1 1 \#$
would represent a beginning of a text consisting of $00$, $\varepsilon$,
$0101$ and $11111$. Furthermore,
each of the words in $\alpha$ would be of length at most $m$.
\item The variable $\beta$ is the current
configuration of a computation to determine $g(2^{m+1} \cdot (m+1))$
(in unary);
this configuration is updated whenever the length and time constraints permit
and the next configuration is shorter than $0^n$, until the
computation finishes.
\item The variable $\gamma$ is a configuration of $M$, while processing the
initial part $\alpha$ of a text for the input language;
note that this configuration includes the memory of $M$ and the portion
of $\alpha$ it has read.
In each cycle this configuration is updated by one more step of the 
computation,
unless the input $\alpha$ is currently exhausted (that is, $M$ would like
to read a symbol which is not yet there) or the length of the configuration
becomes longer than $0^n$. 
\item The variable $e$ is the last completed conjecture of
$M$ and updated whenever the configuration $\gamma$ of $M$ contains a new
value to be output.
\end{iteMize}
In each cycle, the learner $N$ would archive the current
input $x$ on the work tape at a position near to the current one (that is, the
input position has to be reached in linear time) and $N$ would furthermore
update the values of $\beta,\gamma,e,0^m,0^n$ on Tape $0$ ($\alpha$ is
updated only
during some cycles, see below).
\sp
In order to be able to save all required information on the work Tape $1$,
the tape content is modeled as having four tracks. Usually,
only Track $1$ is used for appending new information at the end of the
tape and Track $4$ is used for making sure that computations of the
variables of Tape $0$ meet the time-bound. 
Tracks $2$ and $3$ are used to store data during cycles when some
special operations are needed to transfer data from Tape $1$ to the memory
$\alpha$ in Tape $0$. Furthermore, initially $m=1$. 
\sp
When $\beta$ shows that the computation of
$g(2^{m+1} \cdot (m+1))$ has terminated, and the observed examples
are so long that $n \geq g(2^{m+1} \cdot (m+1))$ then the learner enters
the phase to do special operations (for next several cycles, as many as needed).
Note that eventually this happens for every value of $m$, as the input
language is infinite (assuming it is from $\CalL$).
In each cycle during this special phase, 
from its current position at the end of Tape $1$ back
to the origin $\boxplus$, $N$ will transfer/copy all stored words in 
Tape $1$ of length at most
$m$, which are not already in $\alpha$, to $\alpha$.
During this process, the older words stored in Tracks $2$ and $3$ may be 
erased (but not lost, as they have already been copied to $\alpha$, as each
of them are of length at most $m$). The new input words received during
this phase are copied in Tracks $2$ and $3$ (see below).
Note that a concatenation of
all words up to length $m$ is at most $2^{m+1} \cdot (m+1)$ long (including
separating symbols) and hence $|\alpha| \leq 2^{m+1} \cdot (m+1)$ whenever
$\alpha$ consists only of copies of words up to length $m$ appearing in the
language to be learnt and each such word appears at most once in $\alpha$.
\sp
Now, it is described how special operations are done in the special phases,
see also Figure~\ref{figuretrackexplain} for a rough summary of the
handling of old and new data in each cycle.
When going back on Tape $1$, $N$ will do the following for all words 
$w$ archived in Tracks $1$, $2$, $3$ starting from the current position up to 
$|x|+1$ positions left of the current position (here 
one also considers $w$ that might only partially
overlap with the cells in positions between the current position and
$|x|+1$ to the left of the current position; recall that $x$ is the current
input data to the learner): 
if $|w| \leq m$ then $w$ is compared
with all words in $\alpha$ and in the case that it does not coincide with
any archived word in $\alpha$, $w\#$ is appended at the end of
$\alpha$; note that
all words archived in the Tracks $2$ and $3$ have at most the length $m$.
For each word $w$, this operation needs time $O(|\alpha| \cdot |w|)$.
Note that $w$ has at most length $m$ and $\alpha$ at most length
$2^{m+1} \cdot (m+1)$, giving an overall bound of
$O(2^{m+1} \cdot (m+1) \cdot |w|)$ for the processing of each word $w$.
Furthermore, the concatenation
of all these words archived one after another has length at most $3|x|+3m$;
so one can conclude that the whole operation needs time
$O(2^{m+1} \cdot (m+1) \cdot n)$ which is $O(f(n) \cdot n)$ as
$g(2^{m+1} \cdot (m+1)) \leq n$. Furthermore, all $w$ in Tracks $2$
and $3$ overlapping with the space between the current position in Tape $1$
and the cell at position $|x|$ left of the current position
before the start of the cycle are cleared away as these $w$ all have
at most the length $m$. After the clearance, $x$ will be archived in
Track $2$ (where a special symbol outside the alphabet used for the
archivation data is used to fill up blank spaces, if needed)
and the current position moves by $|x|+1$ to the left.
This is done until the origin $\boxplus$ is reached.
At this point, Track $3$ is empty and
can be used to archive the incoming data in a similar
way while the Turing machine moves back from $\boxplus$ to end of used
part of Tape $1$.
When returning to the usual archivation
mode, $m$ is updated to be the current value of $n$ so that all words
archived in Tracks $2$ and $3$ are again having at most length $m$.
From then onwards, one waits until so much data has been observed
such that the computation of $g(2^{m+1} \cdot (m+1))$ has terminated
and gives a value below (the new value of) $n$.
\sp
One can see from this description 
that, when learning an infinite language,
eventually all words observed will be appended to $\alpha$ and
$M$ will be simulated on the resulting one-one text of the language to
be learnt. Thus, $M$ will eventually stabilise on some index $e$, which will
be taken over as output when the corresponding computation has terminated
and $n$ is larger than $|e|$. This shows that $N$ follows the simulated
learner $M$ and therefore $N$ learns the class to be learnt.\niceqed

\begin{figure}[t]
\begin{center}
\begin{tabular}{|c|c|c|c|} \hline
Mode     & Usual & Backward Special & Forward Special \\ \hline
Old Data Before & ---   & In Tracks 1, 2 and 3 & In Tracks 1 and 2 \\ \hline
Old Data After  & ---   & Into Base Tape and Track 1 &
  Remains unchanged \\ \hline
New Data        & Into Track 1 & Into Track 2 & Into Track 3 \\ \hline
\end{tabular}
\end{center}
\caption{Handling of data at head position of Tape 1. In backward special
mode, old short data is recorded into 
the base tape and old long data remains in Track 1.}
\label{figuretrackexplain}
\end{figure}

\medskip
\noindent
Pitt's original result \cite{Pi89} on linear time learners did not measure the
time in the size of the largest example seen so far, but in the size
of the overall amount of examples seen so far. So the next two results
deal with the question of the additional learning power provided by
one work tape or one stack when the learner can use a Tape $0$ of
length $n$ and run in time linear in $n$ where $n$ is logarithm of the number
of data seen so far plus the length of the longest example seen so far;
hence $n$ increases, though slowly, when a datum is presented multiply.
\sp
Note that in the proof of Theorem~\ref{th:onetapesuperlinear}, the main reason
to use infinite languages and strings of larger and larger length $n$, was to be
able to transfer all stored data of length $m$ onto $\alpha$.
This can also be done if instead of the length $n$, the unbounded growing number
of examples seen so far is used as a parameter to allow the time needed to
do the transfer (in which case additionally, one can make $\alpha$ a fat text). 
For this, one needs to keep track of some earlier maximal length $m'$ and 
number of items $n'$ (including counting the multiple copies, in case they 
are there) so that $2 \cdot m'\cdot n'$ bounds the overall length of all
examples stored in Tracks $2$ and $3$.
When the number of examples seen so far, $n$, is larger than the current
length of $\alpha$ plus $2 \cdot m' \cdot n'$, one
can then start going back, copying new data in Track $2$ until one reaches
the point where the earlier data in Tracks $2$ and $3$ are stored.
At this point one moves all data in Tracks $2$ and $3$ to the end of $\alpha$
which is stored in Tape $0$ and then
starts moving forward on Tape $1$ again,
copying new data into Track $3$ until one reaches
the end of recorded part of all the tracks. At this point one can consider
Track $1$ and Track $2$ as old recorded data (earlier roles played by Tracks $2$
and $3$) and continue recording data in Track $3$ up to the point when
the learner has seen enough examples so as to copy the data in Tracks $1$
and $2$ to Tape $0$. Continuing in this way, one can copy all data to $\alpha$
in Tape $0$ eventually and use the data in Tape $0$ to simulate an automatic
learner on fat text by cycling through the examples archived in $\alpha$.
\sp
This allows to show the following result; its
proof is similar to Theorem~\ref{th:onetapesuperlinear} and
the details are omitted.

\begin{thm}
Let $n$ be the logarithm of the number
of data seen so far plus the length of the longest example seen so far
and consider a learner which can store in Tape $0$ information of length $n$
and can access one additional work tape, with 
update time in each cycle being linear in the corresponding $n$.
Then such a learner can learn every 
learnable automatic family.
\end{thm}

\noindent
The previous and the next result compute the parameter $n$
of the update time and length of Tape $0$ in the same way. While the previous
result showed that one additional work tape is sufficient for full
learning power
under the corresponding linear time model, the next result shows that one
additional stack is insufficient for full learning power.

\begin{thm}
Let $n$ be the logarithm of the number
of data seen so far plus the length of the longest example seen so far
and consider a learner which can store in Tape $0$ information of length $n$
and can access one additional stack, with
update time in each cycle being linear in the corresponding $n$.
Then such a learner fails to learn the class $\CalL$
of all set $L_e = \{0,1\}^*-\{e\}$ where the indices $e$
range over $\{0,1\}^*$.
\end{thm}

\proof
Assume by way of contradiction that such a learner $M$ for $\CalL$
exists. 
\sp
Intuitively, the idea of the proof is that if the learner gets
complex strings (relative to the position), 
then it has to store it in the stack. Thus, if it gets complex strings
in odd positions of the text, and even positions of the text
are filled with simple strings (to form a complete text for some 
target language), then
the learner has to push (codings of) the complex strings on the stack and 
is not able to look at these pushed symbols in later computation.
This allows to construct two such texts for different languages in
the class on which eventually the learner behaves in the same way
(see Claim~\ref{clm-last5}, and then the arguments after this claim);
thus the learner can learn at most one of these two sets.
Claims~\ref{clm-last1} to~\ref{clm-last4} are combinatorial claims 
based on Kolmogorov complexity, needed for proving Claim~\ref{clm-last5}.
Now the formal proof is given.
\sp
Let $\bin(m)$ denote the binary representation of $m$ using $\log(m+2)$ bits
where, for $k \geq 1$,  $\log(k)$ is the downrounded logarithm of base
$2$, that is, the
maximal integer $h$ with $2^h \leq k$.

\begin{clm}\label{clm-last1}
Suppose $\sigma$ and $\tau$ are two finite sequences over $\{0,1\}^*$ such that
$\range(\sigma)-\range(\tau) \neq \emptyset$,
$\range(\tau) - \range(\sigma) \neq \emptyset$,
and $M$ has the same Tape $0$ content and stack content 
after processing either $\sigma$ or $\tau$. Then, $M$ does not learn $\CalL$.
\end{clm}

\noindent
To show the above claim, let $w \in \range(\sigma)-\range(\tau)$
and $w' \in \range(\tau)-\range(\sigma)$.
Let $T'$ be a text for $\{0,1\}^*-\{w,w'\}$. Then,
$M$ has the same convergence behaviour (that is it either diverges or converges
to the same conjecture) on texts $\sigma T'$ and $\tau T'$, which are
texts for $L_{w'}$ and $L_w$ respectively. Thus, $M$ fails to learn at least
one of these languages and thus fails to learn $\CalL$.
This completes the proof of the claim.
\sp
For any recursive text $T$ of any language
satisfying $|T(m)| \leq \log(m+2)$ for all $m$,
define $R_T$ (using an oracle for the halting problem $K$) as follows:
Let $R_T(2m)=T(m)$
and $R_T(2m+1)$ be the string $x$ of length $16(\log(m+2))$ ending
with $\bin(m) 1 0^{\log(m+2)-1}$ which maximises $C(x |
R_T(0)\#R_T(1)\#\ldots\#RT(2m))$.

\begin{clm}\label{clm-last2}
Let $T$ be a recursive text satisfying $|T(m)| \leq \log(m+2)$ for all $m$.
Then the following statements hold:
\begin{enumerate}[(a)]
\item[(a)] $R_T(2j+1)$ are pairwise distinct for different $j$;
\item[(b)] For each $m$, $R_T(2m+1) \notin \{T(i): i \leq m^2\}$;
\item[(c)] For each $m$, $C(R_T(2m+1) | 
       R_T(0)\#R_T(1)\#\ldots\#R_T(2m)) \geq 14\log(m+2)$.
\end{enumerate}
\end{clm}

\noindent
Part (a) follows by definition. Part (b), follows by definition
of $R_T(2m+1)$ and the fact that $16\log(m+2) > \log(m^2+2)$.
For part (c) note that
there exists a string $x$ of length $16 \log(m+2)$, which
ends in $\bin(m)10^{\log(m+2)-1}$, with Kolmogorov complexity
(given $R_T(0)\#\linebreak[3]
R_T(1)\#\linebreak[3] \ldots\linebreak[3] \#R_T(2m)$)
at least $14 \log(m+2)$. As $R_T(2m+1)$ is most complex such string
$x$, part (c) follows.

\begin{clm}\label{clm-last3}
There exists a constant $c_2$ such that the following holds for $m \geq c_2$.
Suppose $T$ is a recursive text satisfying $|T(i)| \leq \log(i+2)$ for all~$i$.
Then,
$C(R_T(2m+1)\#R_T(2m+2)\linebreak[3]
\#\ldots\#R_T(2m+2k-1)| \linebreak[3] 
R_T(0)\#\linebreak[3] R_T(1)\#\linebreak[3] \ldots
\linebreak[3]\#R_T(2m)) \geq k \cdot \log(m+2)$.
\end{clm}

\noindent
To see that the claim holds,
note that for some constant $c_1$, for all $x,y \in \{0,1\}^*$,
$\sigma \in \{0,1,\#\}^*$,
$C((x,y)| \sigma) \geq  C(x | \sigma)+C(y| \sigma\#x)-c_1$,
see \cite{LV08}.
Thus, for all large enough $m$,
\[\eqalign{&C(R_T(2m+1)\linebreak[3] \#R_T(2m+2)\linebreak[3] \#\ldots\#R_T(2m+2k)|
R_T(0)\#R_T(1)\#\ldots\#R_T(2m))\cr
&\geq \sum_{i=m}^{i=m+k-1} [14\log(i+2)-c_1]
 \geq k \log(m+2)}
\] (where the second last inequality follows from
Claim~\ref{clm-last2}(c)).

Let $U^T_i$ and $V^T_i$ denote the Tape $0$ content and stack content of $M$
after processing $R_T(0),\linebreak[3] R_T(1),\linebreak[3] \ldots,
\linebreak[3] R_T(2i)$.

\begin{clm}\label{clm-last4}
There exists a constant $c_3$ such that, for $m$ and $k$ greater than $c_3$,
with $m+k \leq m^2$, the following holds.
Suppose $T$ is a recursive text
satisfying $|T(i)| \leq \log(i+2)$ for all~$i$. Furthermore suppose
that $T$ is computed by a program of Kolmogorov complexity less than 
$6 \log(m+2)$. Then,

$$C((U^T_{m+k},V^T_{m+k}) | (U^T_m,\linebreak[3] V^T_m,k))
\geq \frac{k\log(m+2)}{3}.$$
\end{clm}

\noindent
To show that the claim holds, 
suppose $m$ is large enough as required for Claim~\ref{clm-last3}.
Suppose $C(U^T_{m+k},V^T_{m+k} | U^T_m,V^T_m,k) < \frac{k \log(m+2)}{3}$.
Note that by Claim~\ref{clm-last2}~(b), for $i$ with $m \leq i < m+k$,
$R_T(2i+1)$ does not belong to $T(0), T(1),\ldots, T(m+k)$.
Now, given a program for $T$ and $U^T_m,V^T_m,k,U^T_{m+k},V^T_{m+k}$,
one can construct $w_{2m+1},w_{2m+3},\ldots,w_{2m+2k-1}$ such that,
for $i$ with $m \leq i < m+k$,
\begin{enumerate}[(i)]
\item[(i)] $w_{2i+1}$ ends in $bin(i)10^{\log(i+2)}$,
\item[(ii)] $w_{2i+1} \notin \{T(0), T(1)\ldots T(m+k)\}$,
\item[(iii)] $M$ starting with Tape $0$ content $U^T_m$ and stack
content $V^T_m$, on
input sequence $w_{2m+1}T(m+2) \linebreak[3]
w_{2m+3}\ldots w_{2m+2k-1}T(m+k)$,
ends in  Tape $0$ content being $U^T_{m+k}$ and stack content $V^T_{m+k}$.
\end{enumerate}
Note that $U^T_m$ and $V^T_m$ can be computed using
$R_T(0)\# R_T(1)\# \ldots \#R_T(2m)$. Thus, the expression
\begin{quote}
$C(w_{2m+1}\linebreak[3]\#R_T(2m+2)\#w_{2m+3}\#R_T(2m+4)\ldots\#w_{2m+2k-1} |
\linebreak[3] R_T(0)\# \linebreak[3] R_T(1) \linebreak[3] \# \linebreak[3]
\ldots \linebreak[3] \#R_T(2m))$
\end{quote}
is bounded by
$C(U^T_{m+k},V^T_{m+k}|U^T_m,V^T_m,k)+2\log(k+2)+12\log(m+2)+c' 
\leq \frac{k \log(m+2)}{3}+2\log(k+2)+12\log(m+2)+c'$
for some constant $c'$.
However, by Claim~\ref{clm-last3}, 
\begin{quote}
  $C((R_T(2m+1)\#R_T(2m+2)\#\ldots\#R_T(2m+2k-1))|
  R_T(0)\#R_T(1)\#\ldots\linebreak[3] \#R_T(2m)) \geq k \cdot \log(m+2)$.
\end{quote}
Thus, for large enough $m$,
there are some $h,w,w'$ satisfying $m \leq h < m+k$,
$w' = w_{2h+1} \neq w = R_T(2h+1)$ and
$w,w' \notin \{T(0),T(1),\ldots,T(m+k)\}$. But, then
by Claim~\ref{clm-last1} and Claim~\ref{clm-last2}(a), 
$M$ does not learn $\CalL$.
Hence, $C((U^T_{m+k},V^T_{m+k})|(U^T_k,V^T_k,k)) \linebreak[3]
\geq \frac{k \cdot \log(m+2)}{3}$.
This proves Claim~\ref{clm-last4}.

\begin{clm}\label{clm-last5}
There exists a constant $c_5$ such that
for large enough $m$ and $m+k \leq m^2$ the following holds. 
Suppose $T$ is a recursive text 
satisfying $|T(i)| \leq \log(i+2)$ for all~$i$, and $T$
is computed by a program of Kolmogorov complexity less than $6 \log(m+2)$.
Then, while processing $R_T(2m+1)\linebreak[3] R_T(2m+2)\ldots R_T(2m+2k)$,
\begin{enumerate}[(a)]
\item[(a)] the part of stack consisting of $V^T_m$, except for the top 
$c_5\log(m+2)$ symbols, is never removed and
\item[(b)] $|V^T_{m+k}| \geq (|V^T_m| - c_5\log(m+2))+\frac{k\log(m+2)}{4}$.
\end{enumerate}
\end{clm}

\noindent
To show the claim, consider $k \leq m^2-m$.
Now, $|U^T_{k+m}| = O(\log(k+m+2))$, and 
thus $C(U^T_{k+m}|U^T_m,V^T_m,k) =O(\log(k+m+2))$.
Furthermore, if $V^T_{k+m}=vw$, for some longest prefix $v$ of $V^T_m$, then
the length of the deleted portion of $V^T_m$ (that is $|V^T_k|-|v|$),
can be at most $k \cdot \log(k+m+2) \cdot c'$, for a constant $c'$;
this can be coded using $\log(k+2) + \log\log(k+m+2) + c'$ bits.
Hence, by Claim~\ref{clm-last4}, for some constant $c_4$, $|w| \geq
\frac{k \cdot \log(m+2)}{3} - c_4[\log(m+2) + \log(k+2)]
\geq \frac{k \cdot \log(m+2)}{4}$, for $m, k\geq c''$, for some constant $c''$.
Thus, for large enough $k$, the length of $w$ above 
is larger than what can be removed from the stack in one cycle.
It follows that, for some constant $c_5$, 
the machine $M$, on input $R_T(2m+1),R_T(2m+2),\ldots,R_T(2m+k)$,
does not remove symbols from $V_m^T$,
except maybe for up to $c_5 \log(m+2)$ symbols from the top.
This proves part (a). Part (b) follows, by using the length of $w$
above. This completes the proof of Claim~\ref{clm-last5}.
\sp
Using (a) and (b) of the above claim, it follows that
for large enough $m$,
for each recursive text $T$ of some subset of $\{0,1\}^*$ and
$T$ having a program shorter than $6\log(m+2)$, 
$M$ on $R_T(0), \linebreak[3]
R_T(1),\ldots,R_T(2m)$ will, when processing subsequent data
from $R_T$, never remove symbols from $V_m^T$ except maybe for the
top $c_5 \cdot \log(m+2)$ symbols.

Now, given a program for $T$, $m$,
$R_T(0)\#R_T(1)\# \ldots\# R_T(2m)$,
$U^T_{2m}$ and the topmost $c_5 \cdot \log(2m+2)$ symbols
of $V^T_{2m}$, one can compute a sequence $\sigma$ of length $4m+1$
such that,
\begin{enumerate}[(i)]
\item[(i)] $\sigma(s)=R_T(s)$, for $s \leq 2m$,
\item[(ii)] for all $i \leq 2m$, $\sigma(2i) = T(i)$ and
\item[(iii)] for all $i<2m$, 
   $\sigma(2i+1)$ ends with $\bin(i)10^{\log(i+2)}$ and is of length
   $16 \log (i+2)$.  
\item[(iv)] $M$ after processing
$\sigma$ has Tape $0$ content $U_{2m}^T$ and
the top $c_5 \log(2m+2)$ symbols of the stack are same as the
top $c_5 \log(2m+2)$ symbols of $V_{2m}^T$.
\end{enumerate}
Now, $C(\sigma| R_T(0)\#R_T(1)\#\ldots \# R_T(2m))$, is at most 
$c_6 \cdot \log(m+2)$, for some constant $c_6$,
as it was constructed from a description of $m$ and a description of the top
$c_5 \log(2m+2)$ stack symbols and $O(\log(2m+2))$ symbols of $U_{2m}^T$.
On the other hand, 
$C(R_T(2m+1)\#R_T(2m+2)\# \ldots \#R_T(2m+2m)
 | R_T(0)\#R_T(1)\#\ldots \# R_T(2m))
\geq m \cdot \log(m+2)$ by Claim~\ref{clm-last3}.
\sp
Hence, fix a text $T=T_0$ of the nonempty strings which repeats each
string infinitely often
and let $m$ be large enough and let $\sigma$ be computed as above.
It follows that $\sigma$ and $R_{T_0}(0)R_{T_0}(1)\ldots R_{T_0}(4m)$ 
differ for an $i$ with $m \leq i < 2m$, that is, satisfy
$\sigma(2i+1) \neq R_{T_0}(2i+1)$. Let $w = \sigma(2i+1)$ and
$w' = R_{T_0}(2i+1)$. The strings $w,w'$ do not occur in $T_0(0) T_0(1)\ldots
T_0(m^2)$. Let $m'$ and $m''$ be least such that $T_0(m')=w$ and $T_0(m'')=w'$.
Without loss of generality assume $m'<m''$.
Let $T_1$ and $T_2$ be obtained from $T_0$ as follows:
\begin{iteMize}{$\bullet$}
\item If $T_0(i) = T_0(m')$ then 
$T_1(i) = \varepsilon$ else $T_1(i) = T_0(i)$;
\item If $T_1(i)=T_1(m'')$ then
$T_2(i) = \varepsilon$ else $T_2(i) = T_1(i)$.
\end{iteMize}
Furthermore, the index of $T_1$ has Kolmogorov complexity bounded by
$\log(m'+2)$ and the index of $T_2$ has Kolmogorov complexity bounded
by $2\log(m''+2)$ up to an additive constant. 
When considering $m$ (and thus $m'$ and $m''$) large 
enough, one can absorb this constant into $\log(m'+2)$ and $\log(m''+2)$
respectively, and thus
Kolmogorov complexity of $T_1$ and $T_2$ are bounded by $3\log(m''+2)$.
Now $R_{T_1}$ coincides with $R_{T_0}$ below $2m'$, and
$R_{T_2}$ coincides with $R_{T_1}$ below $2m''$.

In the various claims above (Claim~\ref{clm-last4}, Claim~\ref{clm-last5}), 
when using complexity of $T$ being $6\log(m+2)$, 
only the initial portion of text $T$ of length at most $m^2$ was used.
Thus, it was enough to have the complexity of some text $T'$ coinciding
with $T$ up to first $m^2$ elements having a complexity below $6\log(m+2)$.
Hence, $T_1$ and $T_2$ satisfy the requirements needed in the claims.
\sp
Thus, one can conclude that, when $M$ processes text $R_{T_2}$,
for large enough $s$, the machine $M$ (after having seen the first
$2s+1$ elements of $R_{T_2}$) does not remove more than 
$c_5 \cdot \log(s+2)$
symbols from the top of the stack $V^{T_2}_s$.
Furthermore, if one now replaces the first $4m+1$ members of $R_{T_2}$
by the corresponding members of $\sigma$, then one gets that $M$ on
this new text $R'_{T_2}$ has the same convergence behaviour as on
$R_{T_2}$; however, one text is for $L_w$ while the other one is for
$L_{w'}$, thus these are texts for two different languages and so $M$
does not learn at least one of these languages.\niceqed

\section{Conclusion}

\noindent
The starting point of this research is that automatic functions can be
characterised using one-tape Turing machines. More precisely, a function
is automatic iff it is computed by a position-faithful one-tape Turing
machine in linear time. This is the smallest reasonable linear time
complexity class and so the automatic functions turn out to sit at the
bottom of the corresponding hierarchy. An open problem is whether the
corresponding formalisation using alternating linear time position-faithful
one-tape Turing machines also characterises the automatic functions.
\sp
Automatic functions have been investigated in learning theory in order
to model resource-bounded learners. Due to Pitt's delaying trick \cite{Pi89},
unrestricted recursive learners can be bounded heavily in the time
that they use without
losing learning power. However,
automatic learners are not able to learn every learnable class, as their
ability to memorise data is insufficient. Therefore, one might ask
whether one can replace an automatic learner by a linear-time learner
working on a one-tape Turing machine with a tape of length bounded
by the longest datum seen so far plus some additional memory.
\sp
These additional memory devices are not restricted in length, 
though restricted in
the amount of access the learner has per cycle: In each cycle the learner
runs in time linear in the longest example seen so far, updates the
base tape and accesses the additional storage devices only to retrieve
or store a linear number of symbols. It is shown that two additional work
tapes, two additional stacks or one additional queue give full learning
power; furthermore, the learning power of one additional stack is properly
intermediate and the learning power of one additional work tape is better
than no additional work tape. It is an open problem whether there is a
difference in the learning power of one and two additional work tapes.
\sp
For some special cases and slightly superlinear computation time, it
was possible to show that one additional work tape is enough. The methods
of this proof do not generalise to the general case.

\end{document}